\documentclass[longauth]{aa} 

\usepackage{graphicx}
\usepackage{amsmath, amssymb}
\usepackage{caption, subcaption}
\usepackage{geometry}
\usepackage{comment}
\usepackage{tabularx}
\usepackage{fixltx2e}
\usepackage{orcidlink}
\usepackage{natbib}
\usepackage{footnote}
\usepackage{tablefootnote}
\usepackage[utf8]{inputenc}
\usepackage{txfonts}
\usepackage{placeins}
\usepackage{dblfloatfix}
\usepackage{caption} 

\DeclareSymbolFont{CMletters}{OML}{cmm}{m}{it}
\DeclareMathSymbol{\nu}{\mathord}{CMletters}{23}
\DeclareMathSymbol{v}{\mathord}{CMletters}{`v}

\hypersetup{
    colorlinks=true,
    linkcolor=blue,
    citecolor=blue,
    urlcolor=blue
}

\begin{document} 

\title{The ALMA survey to Resolve exoKuiper belt Substructures (ARKS)}
\subtitle{VII: Optically thick gas with broad CO gaussian local line profiles in the HD~121617 disc
  }
\author{A.~Brennan\textsuperscript{1}\orcidlink{0000-0002-7050-0161}\thanks{E-mail: brenna29@tcd.ie}
          \and
          L.~Matr\`a\textsuperscript{1}\orcidlink{0000-0003-4705-3188}
          \and
          S.~Mac~Manamon\textsuperscript{1}\orcidlink{0000-0002-7066-8052}
          \and
          S.~Marino\textsuperscript{2}\orcidlink{0000-0002-5352-2924}
          \and
          G.~Cataldi\textsuperscript{3,4}\orcidlink{
0000-0002-2700-9676}
          \and
          A.~M.~Hughes\textsuperscript{5}\orcidlink{0000-0002-4803-6200}
          \and
          P.~Weber\textsuperscript{6,7,8}\orcidlink{0000-0002-3354-6654}
          \and
          Y.~Han\textsuperscript{9}
          \and
          J.~P.~Marshall\textsuperscript{10}\orcidlink{0000-0001-6208-1801}
          \and
          B.~Zawadzki\textsuperscript{5}\orcidlink{0000-0001-9319-1296}
          \and
          P.~Luppe\textsuperscript{1}\orcidlink{
0000-0002-1018-6203}
          \and
          A.~A.~Sefilian\textsuperscript{11}\orcidlink{0000-0003-4623-1165}
          \and
          A.~Mo\'or\textsuperscript{12}\orcidlink{0009-0001-9360-2670}
          \and
          M.~A. MacGregor\textsuperscript{13}\orcidlink{0000-0001-7891-8143}
          \and
          J.~B.~Lovell\textsuperscript{14}\orcidlink{0000-0002-4248-5443}
          \and
          A.~K\'osp\'al\textsuperscript{12,15}\orcidlink{0000-0001-7157-6275}
          \and 
          M.~Bonduelle\textsuperscript{16}\orcidlink{0009-0000-7049-8439}
          \and
          E.~Mansell\textsuperscript{5}\orcidlink{0009-0004-1433-8149}
          \and
          M.~C.~Wyatt\textsuperscript{17}\orcidlink{0000-0001-9064-5598}
          \and          
          T.~D.~Pearce\textsuperscript{18}\orcidlink{0000-0001-5653-5635}
          \and
          J.~M.~Carpenter\textsuperscript{19}\orcidlink{0000-0003-2251-0602}
          \and
          D.~J.~Wilner\textsuperscript{13}\orcidlink{0000-0003-1526-7587}
          \and
          C.~del~Burgo\textsuperscript{20,21}\orcidlink{0000-0002-8949-5200}
          \and     
          S.~P\'erez\textsuperscript{6,7,8}\orcidlink{0000-0003-2953-755X}
          \and
          Th.~Henning\textsuperscript{15}
          \and
          J.~Milli\textsuperscript{16}\orcidlink{0000-0001-9325-2511}
          \and
          E.~Chiang\textsuperscript{22}\orcidlink{0000-0002-6246-2310}
}

\institute{School of Physics, Trinity College Dublin, the University of Dublin, College Green, Dublin 2, Ireland \and
Department of Physics and Astronomy, University of Exeter, Stocker Road, Exeter EX4 4QL, UK \and
National Astronomical Observatory of Japan, Osawa 2-21-1, Mitaka, Tokyo 181-8588, Japan \and
Department of Astronomy, Graduate School of Science, The University of Tokyo, Tokyo 113-0033, Japan \and
Department of Astronomy, Van Vleck Observatory, Wesleyan University, 96 Foss Hill Dr., Middletown, CT, 06459, USA \and
Departamento de Física, Universidad de Santiago de Chile, Av. V\'ictor Jara 3493, Santiago, Chile \and
Millennium Nucleus on Young Exoplanets and their Moons (YEMS), Chile \and
Center for Interdisciplinary Research in Astrophysics Space Exploration (CIRAS), Universidad de Santiago, Chile \and
Division of Geological and Planetary Sciences, California Institute of Technology, 1200 E. California Blvd., Pasadena, CA 91125, USA \and
Academia Sinica Institute of Astronomy and Astrophysics, 11F of AS/NTU Astronomy-Mathematics Building, No.1, Sect. 4, Roosevelt Rd, Taipei 106319, Taiwan \and
Department of Astronomy and Steward Observatory, University of Arizona, Tucson, AZ 85721, USA \and
Konkoly Observatory, HUN-REN Research Centre for Astronomy and Earth Sciences, MTA Centre of Excellence, Konkoly-Thege Mikl\'os \'ut 15-17, 1121 Budapest, Hungary \and
Department of Physics and Astronomy, Johns Hopkins University, 3400 N Charles Street, Baltimore, MD 21218, USA \and
Center for Astrophysics | Harvard \& Smithsonian, 60 Garden St, Cambridge, MA 02138, USA \and
Max-Planck-Insitut f\"ur Astronomie, K\"onigstuhl 17, 69117 Heidelberg, Germany \and
Univ. Grenoble Alpes, CNRS, IPAG, F-38000 Grenoble, France \and
Institute of Astronomy, University of Cambridge, Madingley Road, Cambridge CB3 0HA, UK \and
Department of Physics, University of Warwick, Gibbet Hill Road, Coventry CV4 7AL, UK \and
Joint ALMA Observatory, Avenida Alonso de C\\'ordova 3107, Vitacura 7630355, Santiago, Chile \and
Instituto de Astrof\'isica de Canarias, Vía L\'actea S/N, La Laguna, E-38200, Tenerife, Spain \and
Departamento de Astrof\'sica, Universidad de La Laguna, La Laguna, E-38200, Tenerife, Spain \and
Department of Astronomy, University of California, Berkeley,
Berkeley, CA 94720-3411, USA}

\date{Received x, 2025; accepted x, 2025}

\abstract{CO gas has been detected in $\sim$20 debris discs, typically classified as CO-poor or CO-rich. We present observations of the CO-rich HD~121617 debris disc as part of the ALMA survey to Resolve exoKuiper belt Substructures (ARKS).}
{We model local CO line profiles in the HD~121617 debris disc to investigate optical depth, CO mass, and temperature. HD~121617 is a well-suited ARKS target due to its previously detected CO emission and moderate inclination, reducing the effect of Keplerian shear.}
{Using high-resolution ALMA Band 7 observations of $^{12}$CO~J=3-2 (26 m s$^{-1}$, $0\farcs1$), we create local line profiles by aligning and stacking spectra in concentric annuli of $0\farcs02$ width. These profiles are modelled with both a toy model and a \texttt{RADMC-3D} model that includes projection effects and Keplerian shear.}
{The resulting local profiles are Gaussian-shaped and broad due to the effect of Keplerian shear. Fitting a \texttt{RADMC-3D} model to the $^{13}$CO data, we find that an optically thick model (temperature of 38 K and mass of $2 \times 10^{-3}$ M$_{\oplus}$) reproduces the data, particularly the enhanced intensity at orbital azimuths of $\sim$ $\pm45^{\circ}$ and $\pm135^{\circ}$, which forms an X-shape in the velocity integrated intensity map, as well as the broader $^{12}$CO linewidth compared to $^{13}$CO. Scaling this model by the ISM abundance ratio ($\sim$77) also reproduces the $^{12}$CO data, but high optical depths and model assumptions limit mass constraints.}
{Keplerian shear causes azimuthally averaged line profiles to appear Gaussian regardless of optical depth; therefore, we caution against using the local line profiles to distinguish between optically thin and thick emission. We constrain the mean molecular weight to $12.6_{-1.1}^{+1.3}$, dependent on model assumptions. Although model dependent, our $^{13}\textrm{CO}$ results indicate that C$^{18}$O might also be optically thick in CO-rich debris discs, contrary to previous assumptions, and that the mean molecular weight is significantly higher than if H$_2$ were the dominant gas species, suggesting a non-primordial composition.}
  
               
\titlerunning{Broad CO profiles in the HD~121617 exoKuiper belt}
\maketitle

\section{Introduction}

The detection of gas in debris discs where CO is the most commonly detected species challenges current models of planetary system evolution \citep{Hughes(2018)}. These discs generally fall into two categories based on their detected CO mass. Some, such as Fomalhaut, which has a detected CO mass of $\sim10^{-7}$M$_\oplus$ \citep{Matra(2017)} are CO-poor. Others are CO-rich, typically associated with A-type stars, and exhibit CO masses greater than $\sim10^{-4}$M$_\oplus$ \citep{S.Marino(2020)}. However, the origin and evolution of this gas remain unclear. In discs with low CO levels, the gas is thought to be secondary, generated through a collisional cascade. In this process, mutual collisions between dust grains and planetesimals in the disc result in the creation of both gas and dust \citep[e.g.][]{Marino(2016), Marino(2017), L.Matra(2017)}. A secondary gas origin in CO-poor discs is implied by the short photodissociation timescale of CO ($\sim$130 years), due to the interstellar radiation field \citep[ISRF,][]{Heays(2017)}. As a result, any observed CO in these discs must be continuously replenished. In contrast, the origin of CO in CO-rich discs is less certain. Initially, these discs were thought to be hybrid, with the observed CO gas being a remnant of the gas-rich protoplanetary phase and the dust being of secondary origin \citep{Kospal(2013)}. In this scenario, high CO levels are shielded by undetected molecular hydrogen (H$_{2}$) left over from the protoplanetary stage. More recently, the \ion{C}{I} shielding scenario has suggested that some CO-rich exoKuiper belts may have a secondary gas origin \citep[][]{Q.Kral(2019)}. In this scenario, the photodissociation of CO produces \ion{C}{I} that shields the remaining CO, slowing photodissociation and allowing CO to accumulate through the collisional cascade. 

Accurately determining the CO mass is crucial for inferring a gas origin. Determining the CO mass in exoKuiper belts is relatively straightforward in theory for discs with low CO line luminosities, as multiple CO transitions are expected to be optically thin when observed with the Atacama Large Millimeter/submillimeter Array \citep[ALMA,][]{L.Matra(2017)}. In such cases, CO mass can be directly measured from the optically thin emission lines \citep[e.g.][]{Kospal(2013), Matra(2017)}. However, the process is more difficult for CO-rich discs. In CO-rich discs, $^{12}$CO is typically assumed to be optically thick due to the low observed $^{12}$CO/$^{13}$CO and $^{12}$CO/C$^{18}$O flux ratios. As a result, the CO mass is inferred from measured isotopologue line fluxes, typically of C$^{18}$O \citep[e.g.][]{Moor(2019)}, and then re-scaled to the total CO mass assuming interstellar medium (ISM) abundance ratios \citep[][]{Wilson(1994)}. However, this assumption may not hold universally. For example, a recent study of the protoplanetary disc TW Hya reported a $^{12}$CO/$^{13}$CO ratio of $21 \pm 5$ at disc radii of 70-110 au \citep[][]{Yoshida(2022)}, which is significantly lower than the ISM ratio of $\sim77$ \citep[][]{Wilson(1994)}. This result challenges the assumption that ISM ratios can be used to determine CO masses in protoplanetary discs \citep{Miotello(2014)}. In debris discs, the same could be true depending on the gas origin, CO formation, and CO dissociation processes.

The ALMA observations analysed in this paper were obtained as part of the first ALMA large programme dedicated to exoKuiper belts: the ALMA survey to Resolve exoKuiper belt Substructures (ARKS). ARKS observed 18 exoKuiper belts at high resolution and combined these with archival observations of six additional systems, resulting in a complete sample of 24 targets that enables a detailed investigation of both the gas and dust components in exoKuiper belts. An overview of the ARKS programme is provided in \citet{Seba_arks}.

In this work, we used high spectrospatial resolution $^{12}\textrm{CO}$ and $^{13}\textrm{CO}$ observations of a moderately inclined CO-rich exoKuiper belt to investigate the local spectral line profiles, in the context of its optical depth, for the first time. High spectral resolution is essential for accurately resolving the line profile of CO emission across the disc, with high spatial resolution minimising the contribution of Keplerian shear to the local linewidth. The effect of Keplerian shear causes the line profile to broaden due to variations in gas velocities within a single beam; therefore, reducing the beam size can significantly reduce its broadening effect. Additionally, the effect of Keplerian shear is reduced by selecting a moderately inclined exoKuiper belt, such as HD~121617, the subject of this study, which has an inclination of 41 $\pm$ 0.6$^\circ$ \citep{Seba_arks}. The moderate inclination of the disc causes the range of projected Keplerian velocities probed by a given line-of-sight to be narrower and thus reduces the contribution of Keplerian shear to the observed line profile.

HD~121617 is an A1V main-sequence star \citep[][]{Houk(1978)} and a member of the Upper Centaurus Lupus (UCL) association \citep[][]{Hoogerwerf(2000)}. The star has an estimated age of 16~$\pm$~2~Myr, based on the age of the UCL association \citep[][]{Pecaut(2016)} at a distance of 117.9~$\pm$~0.5~pc \citep[][]{Gaia(2023)}. HD~121617 hosts an exoKuiper belt in which gas was first detected using ALMA \citep[][]{Moor(2017)}. Previous low-resolution observations showed that the gas disc has an inner radius of $\sim$50 au and an outer radius of $\sim$100 au, with a CO gas mass of 0.02 M$_{\oplus}$ based on assumed optically thin C$^{18}$O transition observations and the assumption of ISM-like isotopologue ratios \citep[][]{Moor(2017)}. Additionally, \citet{Smirnov-Pinchukov(2022)} conducted observations of CN, HCN, HCO$^{+}$, CCH, CS, CO, $^{13}$CO, and C$^{18}$O, with CO being the detected molecule, which potentially suggests a secondary gas origin. In \citet[][]{Clement(2023)}, the first resolved image of the exoKuiper belt in scattered light features a ring with a sharply defined inner edge, possibly shaped by an undiscovered planet or affected by gas drag \citep{Clement(2023)}. 

HD 121617 is a recurring target in the ARKS collaboration. In \citet{Seba_arks}, the following best-fit parameters for the continuum were determined assuming a Gaussian distribution: FWHM = 18 $\pm$ 1 au, PA = 58.7 $\pm$ 0.7$^\circ$ (position angle), dust mass = 0.21~M$_{\oplus}$, a centre offset by $\mathrm{RA} = 8 \pm 8$ mas (right ascension), and $\mathrm{DEC} = 14 \pm 7$ mas (declination) relative to the phase centre. \citet{Yinuo_arks} presents non-parametric radial profiles, suggesting a single narrow belt with a scale height aspect ratio of 0.090$^{+0.010}_{-0.009}$ and 0.071$^{+0.046}_{-0.049}$, fitted with \texttt{frank} \citep{jennings(2020), Terrill(2023)} and \texttt{rave} \citep{Han(2022), Han(2025)}. \citet{Julien_arks} find a clear offset between the polarised scattered light and the ALMA continuum, with CO and the continuum peaking slightly interior to the scattered light. \citet{Josh_arks} report the most significant continuum asymmetry in the ARKS sample: a $\sim50$\% brightness enhancement in the south-west ansa, forming an arc spanning $\sim$90$^\circ$. \citet{Seba_2arks} reveal non-Keplerian gas kinematics, interpreted as a pressure gradient from gas concentrated in a narrow belt. \citet{Philipp_arks} explore gas drag effects on dust dynamics and the conditions under which the observed continuum asymmetry could arise from dust trapping in a gas vortex. Another explanation is that the dust clump is driven by interactions with an unseen planet \citep{Tim}. In this scenario, an outwards-migrating planet sweeps material into mean-motion resonances, which produces the clump structure. A similar model was previously proposed to explain tentative dust clumps in the $\epsilon$ Eri debris disc \citep{Booth(2023)}.

The remainder of this paper is organised as follows: In Section \ref{sec:Observations}, we describe the ALMA observations. In Section \ref{sec:ObservationsResults}, the observational results, including velocity integrated intensity, velocity, and peak intensity maps, as well as radial profiles, are presented. In Section \ref{sec:Local Spectral Line Profiles}, we detail the creation of local line profiles and introduce the simple radiative transfer toy model used to fit these profiles. In Section \ref{sec:RADMC-3D}, we describe our \texttt{RADMC-3D} model, and in Section \ref{sec:Discussion}, we conclude with a discussion of optical depth and CO mass, followed by the conclusions in Section \ref{sec:conclusion}.

\section{Observations}
\label{sec:Observations}

We observed HD~121617 in Band 7 during Cycle 9 as part of the ALMA survey ARKS large programme (project number 2022.1.00338.L, PI: S. Marino). Observations were conducted between October 2022 and May 2023 using the C-3 and C-6 antenna configurations. Full details on the observation setup, calibration, and imaging can be found in \citet{Seba_arks} and \citet{Sorcha_arks}.

The correlator was configured with four spectral windows. Two windows were allocated for observing the dust continuum centred at 344 GHz and 333.2 GHz, with a bandwidth of 1875 MHz and a spectral resolution of 31.256 MHz. The remaining two windows were designated for CO line studies, where we observed both $^{12}\textrm{CO}$ J=3-2 and $^{13}\textrm{CO}$ J=3-2 line emission (hereafter simply $^{12}\textrm{CO}$ and $^{13}\textrm{CO}$). The $^{12}\textrm{CO}$ line was observed within a window centred at 345.796~GHz, a spectral resolution of 30.518~kHz (26~m~s$^{-1}$), and a bandwidth of 58.59~MHz. The $^{13}\textrm{CO}$ line was observed within a window centred at 331.3 GHz, a resolution of 975.563 kHz (884 m s$^{-1}$), and a bandwidth of 1875 MHz.

The calibration and imaging were done using the standard ALMA reduction tool Common Astronomy Software Applications (CASA version 6.4.1.12). In summary, to extract the continuum image and the spectral cubes of different lines from the calibrated visibilities, the \texttt{tclean} CASA task was used with Briggs weighting, and a robust parameter of 0.5. The achieved sensitivities and synthesised beam sizes for each of the $^{12}\textrm{CO}$ and $^{13}\textrm{CO}$ cubes are given in Table \ref{tab:observationaldata}.

\begin{table}
\caption{Cube properties for $^{12}\textrm{CO}$ J=3-2 and $^{13}\textrm{CO}$ J=3-2.} 
\label{tab:observationaldata}
\centering
\begin{tabular}{l c c} 
\hline\hline
                              & $^{12}\textrm{CO}$ & $^{13}\textrm{CO}$ \\ \hline
Spectral resolution (m s$^{-1}$)     & 26          & 884      \\ 
Beam Size (arcsec $\times$ arcsec) & 0.12 $\times$ 0.11 & 0.13 $\times$ 0.12 \\ 
Beam PA (deg)              & -76.5       & -72.6       \\ 
RMS noise level (mJy beam\(^{-1}\)) & \(4.7\) & \(14.8\) \\ 
\hline
\end{tabular}
\tablefoot{Note that the quoted beam sizes and RMS noise levels are specifically for robust = 0.5.}
\end{table}

\section{Observational results}
\label{sec:ObservationsResults}

\begin{figure*}
\centering
  \includegraphics[width=1\linewidth]{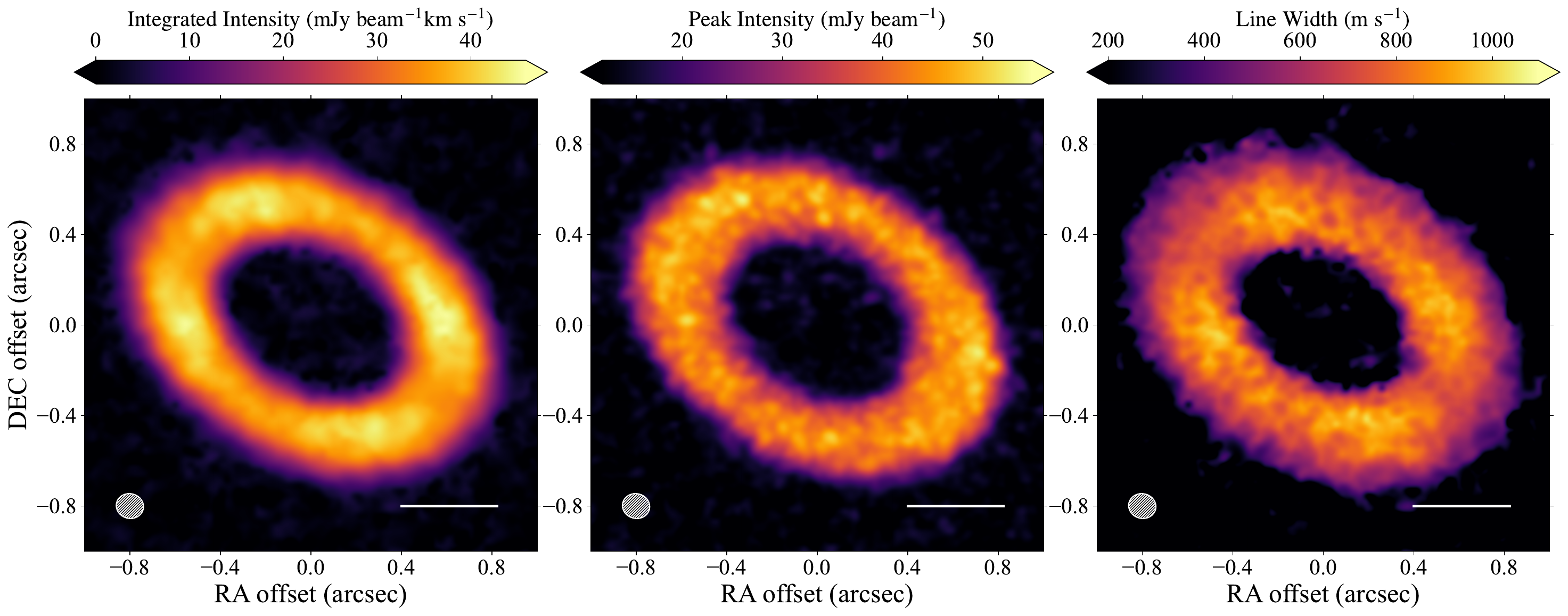}
  \caption{$^{12}\textrm{CO}$ emission for HD~121617. Left: Velocity integrated intensity map created by shifting the spectral emission and integrating along the velocity axis for all channels with emission above 4$\sigma$. Middle: Peak intensity map generated using the quadratic method with \texttt{bettermoments}. Right: Linewidth FWHM map produced by dividing the velocity integrated intensity map by the peak intensity map and converting it to FWHM, assuming a Gaussian line profile shape. The beam size is shown in the lower-left corner of each panel. The horizontal white bar indicates 50 au. The corresponding maps for $^{13}\textrm{CO}$ can be found in Appendix \ref{sec:addional_figures}.}
  \label{fig:12COmmaps}
\end{figure*}

We detect both $^{12}\textrm{CO}$ and $^{13}\textrm{CO}$ emission in HD~121617's exoKuiper belt. Velocity integrated intensity maps for both $^{12}\textrm{CO}$ (Fig. \ref{fig:12COmmaps}, left panel) and $^{13}\textrm{CO}$ (Fig. \ref{fig:13COmmaps}) were generated by shifting the emission to the systemic velocity (7851 m s$^{-1}$), as described in \citet{Sorcha_arks}. The shifted cube was then integrated along the velocity axis for all pixels with peak emission above a 4$\sigma$ threshold. Additionally, we present peak intensity maps for $^{12}\textrm{CO}$ and $^{13}\textrm{CO}$ in Fig. \ref{fig:12COmmaps} (middle panel) and Fig. \ref{fig:13COmmaps_2}, respectively. These peak intensity maps were created using the quadratic method within the \texttt{bettermoments} package \citep[][]{Teague(2018)}. This method identifies the channel with the peak intensity, fits a quadratic curve to the pixel as well as its two neighbouring pixels, and calculates the associated uncertainty. For further details on the creation of velocity integrated intensity and peak intensity maps, see \citet{Sorcha_arks}. Finally, we present $^{12}\textrm{CO}$ linewidth maps in Fig. \ref{fig:12COmmaps} (right panel), generated by dividing the velocity integrated intensity map by the peak intensity map and converting it to FWHM, assuming a Gaussian line profile shape.

In the velocity integrated intensity maps, we observe enhanced intensity at $\pm45^{\circ}$ and $\pm135^{\circ}$ (with $0^{\circ}$ measured from the top-left ansa), which forms an X-shape in the emission map. This azimuthal, spectrally integrated intensity variation could be potentially due to optical depth effects, with this X-shape being even more pronounced in the linewidth map due to the azimuthal dependence of Keplerian shear (see Appendix \ref{sec:keplerianshear} for details). The effect of Keplerian shear arises as gas within a single ALMA beam orbits the star at slightly different line-of-sight projected velocities, depending on its radius and azimuth. For example, gas closer to the star moves faster, while gas farther out moves more slowly, creating a velocity gradient within the beam that broadens the emission. In Appendix \ref{sec:keplerianshear}, we present an analytical derivation showing that, for an inclined disc, the effect of Keplerian shear decreases with radius and consistently produces an azimuthal X-shape pattern in the broadening of a line. Finally, in the peak intensity maps, we note enhanced peak intensity at the ansae.

In \citet{Sorcha_arks}, the integrated line flux is calculated from the spectrospatially stacked spectra, where emission from different pixels was shifted spectrally to correct for their sky-projected Keplerian velocities. This alignment enhanced the signal-to-noise ratio (S/N) by coherently summing the spectra, ensuring that any detected gas appeared as a single peak at the stellar velocity rather than a Keplerian profile. The integrated line flux was then determined by integrating the spectrospatially stacked spectrum, with bootstrapping used to calculate the uncertainty where the integral was calculated multiple times in regions of the spectrospatially stacked spectra that only contained noise. An additional 10\% flux density calibration error was also added in quadrature. For HD~121617, they derived an integrated line flux of 3.6 $\pm$ 0.4 Jy km s$^{-1}$ for $^{12}\textrm{CO}$ and 1.5 $\pm$ 0.2 Jy km s$^{-1}$ for $^{13}\textrm{CO}$ \citep{Sorcha_arks}.

\begin{figure}
\centering
  \includegraphics[width=\linewidth]{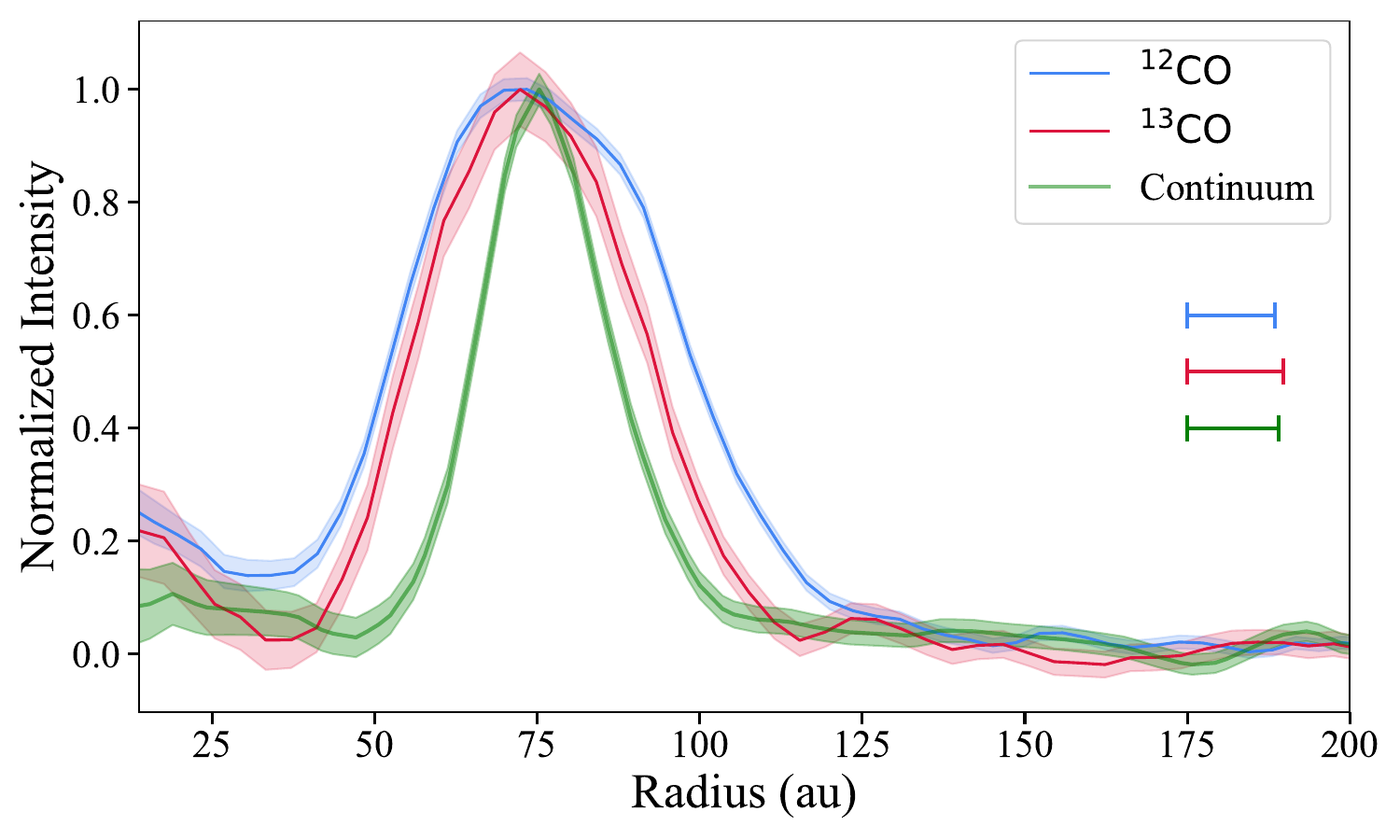}
  \caption{Normalised radial profile for $^{12}\textrm{CO}$ (blue) and $^{13}\textrm{CO}$ (red) produced using \texttt{gofish} and continuum (green) produced using \texttt{frank}. These radial profiles are not de-convolved and, therefore, might not be directly comparable in width. The lower x-axis limit in each radial profile is set at the first resolved resolution element of the $^{12}\textrm{CO}$ observation. The blue, red, and green bars show the resolution of the observations.}
  \label{fig:radial_profile}
\end{figure}

In Fig. \ref{fig:radial_profile}, we present radial profiles for $^{12}$CO, $^{13}$CO, and the continuum. The gas radial profiles were generated using the \texttt{gofish} package\footnote{The \texttt{gofish} Python package is available at \url{https://github.com/richteague/gofish}.}, which divides the disc gas emission into annuli and spectrospatially stacks the spectra within each annulus, assuming a Keplerian rotation pattern for the disc \citep{Teague(2019b)}. Then, we integrated over each annulus's stacked spectrum to report integrated line intensity as a function of radius. The continuum radial profile was produced using \texttt{frank} with the full procedure detailed in \citet{Yinuo_arks}. Additionally, the $^{12}$CO, $^{13}$CO, and continuum profiles were normalised to facilitate the comparison of their relative radial extents. In Fig.\ref{fig:radial_profile}, we observe that the $^{12}\textrm{CO}$ and $^{13}\textrm{CO}$ disc are detected between $\sim$45 au and 125 au. The $^{13}\textrm{CO}$ emission peaks at $\sim$72 au, $^{12}\textrm{CO}$ at $\sim$73 au and the continuum at $\sim$75 au. Furthermore, inwards of this main ring, both the $^{12}\textrm{CO}$ and $^{13}\textrm{CO}$ emissions dip before rising again interior to $\sim$34 au.

A velocity map was created for $^{12}\textrm{CO}$ using \texttt{bettermoments}, where all pixels with peak intensities below 5$\sigma$ are masked. The velocity map was then produced using the Gaussian method, where a Gaussian is fitted to each pixel; and the uncertainty in its peak velocity was derived from the \texttt{bettermoments} fit (see \citet{Sorcha_arks, Seba_2arks} for velocity maps). We proceeded to fit the velocity map with a simple Keplerian model, assuming a geometrically thin disc, where in every pixel
\begin{equation}
v_{0} = \sqrt{\frac{GM_{\star}}{r}} \cos(\phi) \sin(i)+v_{\mathrm{BARY}}
\label{eq:1},
\end{equation}
where $v_{0}$ is the peak line-of-sight velocity, $G$ is the gravitational constant, $M_{\star}$ is the stellar mass, $\phi$ is the relevant azimuth, $i$ is the inclination, and $v_{\mathrm{BARY}}$ is the system velocity. We fixed the distance parameter to the known literature value of 117.9 $\pm$ 0.5 pc \citep[][]{Gaia(2023)}. The other parameters, which are free, are summarised in Table \ref{tab:rotbestfit}. 

To sample the posterior probability distribution of the model parameters, we used an affine-invariant ensemble sampler implemented through the \texttt{emcee} package \citep[][]{Goodman(2010), Foreman-Mackey(2013)}. We employed uniform priors with well-defined boundaries, ensuring they were sufficiently distanced from the best-fit values. We ran the Markov chain Monte Carlo (MCMC) method for 1500 steps with 32 walkers after an additional burn-in period of 500 steps and visually assessed convergence. The best-fit values and associated uncertainties are reported in Table \ref{tab:rotbestfit} as the 50 $\pm$ 34 percentile range of the probability distribution of each parameter, marginalised over all others. The PA, inclination, RA, and DEC offsets are consistent with those measured for the continuum  \citep{Seba_arks}. The stellar mass of 1.89 M$_{\odot}$ is consistent with that derived using stellar models \citep[$1.90\pm0.01\ $M$_{\odot}$,][]{Seba_arks}. 

\begin{table}
\centering
\caption{Best-fit Keplerian model parameters from fitting HD12167's $^{12}\textrm{CO}$ velocity map.}
\label{tab:rotbestfit}
\begin{tabular}{l c}
\hline\hline
                              &    \\ \hline
Position angle (deg)          & 59.6 $\pm$ 0.01     \\ 
Inclination (deg)            & 43.8 $\pm$ 0.02     \\ 
Stellar mass (M$_{\odot}$)           & 1.89 $\pm$ 0.001  \\ 
$v_{\textrm{BARY}}$  (m s$^{-1}$)                   & 7860 $\pm$ 1 \\ 
RA offset (mas)                      & -3.1 $\pm$ 0.1 \\ 
Dec offset (mas)                      & 19.8 $\pm$ 0.1 \\ 
\hline
\end{tabular}
\tablefoot{The quoted best-fit values are the median of the marginalised distribution. The uncertainties are based on the 16th and 84th percentiles of the marginalised distributions. We note that these uncertainties are likely underestimated.}
\end{table}

\citet{Seba_2arks} find that the residuals map after subtracting the best-fit Keplerian model is not perfectly noise-like. Although a Keplerian model describes the bulk of the velocity map well, some deviations from perfectly Keplerian rotation are present. In particular, there is structure hinted at ansae, which is discussed in \citet{Seba_2arks}. The residuals show a pattern that indicates that the gas is sub-Keplerian in the outer regions and super-Keplerian in the inner regions. This is likely caused by (respectively) positive and negative radial pressure gradients towards the ring centre. These gradients would likely correspond to a peak in gas density and therefore pressure \citep{Seba_2arks}.

\section{Local spectral line profiles}
\label{sec:Local Spectral Line Profiles}

\subsection{Local line profile extraction}
\label{sec:Local Line Profile Extraction}

\begin{figure*}
\centering
  \includegraphics[width=0.8\linewidth]{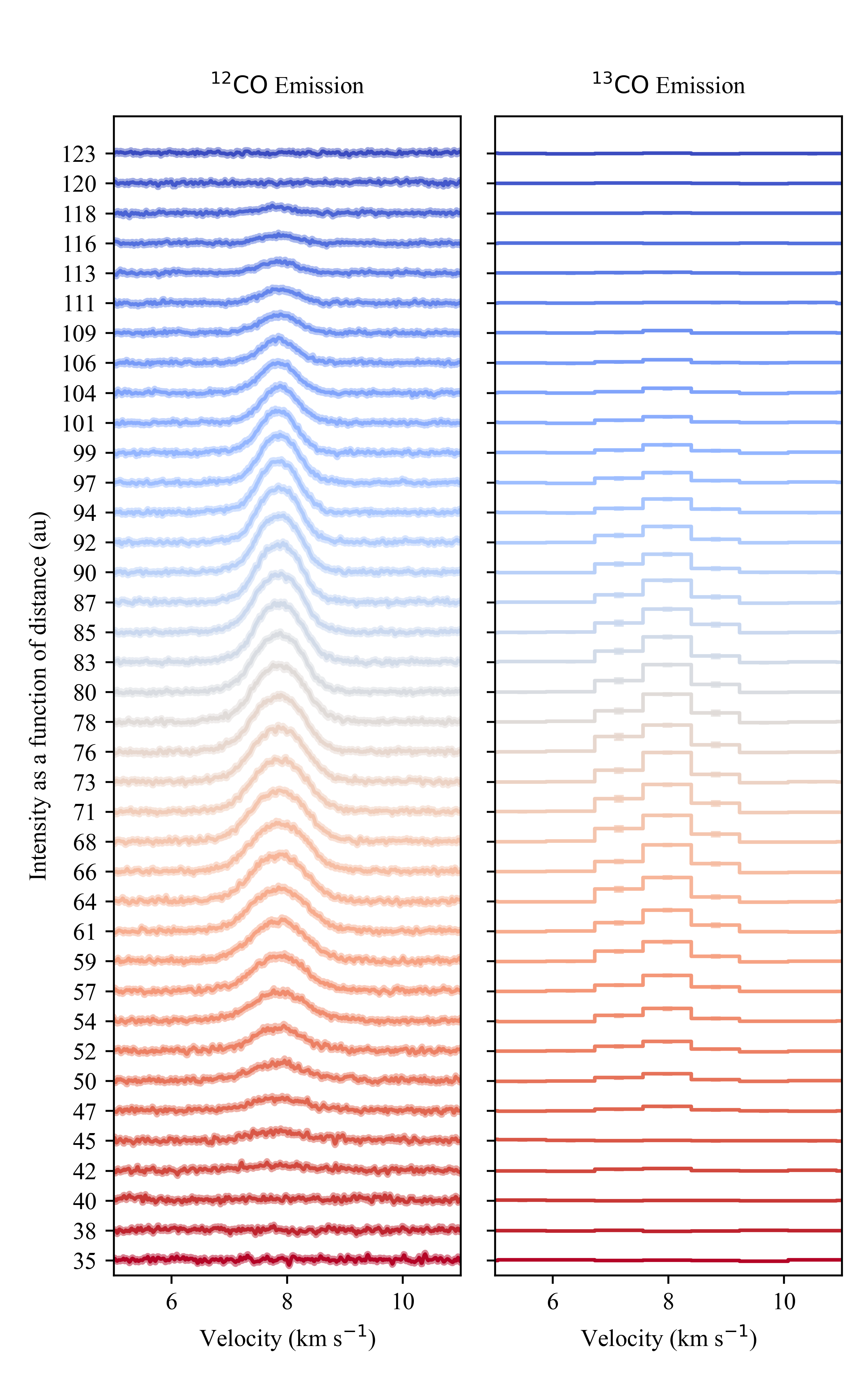}
  \caption{Local line profiles for $^{12}\textrm{CO}$ (left panel) and $^{13}\textrm{CO}$ (right panel) as a function of distance from the star. The error bars for $^{12}\textrm{CO}$ in the left panel (shaded regions) and for $^{13}\textrm{CO}$ in the right panel (error bars) have been re-scaled by constant factors $f_{12}$ and $f_{13}$, respectively, as fitted in Section \ref{sec:1DResults}. These errors are small and only marginally visible.}
  \label{fig:stacked_spectra}
\end{figure*}

To obtain spectral line profiles that are localised to specific radial distances from the star, we aligned and stacked spectra within annuli of $0\farcs02$ width. This width is smaller than the beam size for both $^{12}\textrm{CO}$ and $^{13}\textrm{CO}$ (Table \ref{tab:observationaldata}). By choosing an annulus width smaller than the beam size, we reduced the effect of Keplerian shear by limiting the radial extent of the disc contributing to the emission. This ensured that the emission within each annulus corresponds to roughly one beam width, within which the pixels are correlated. We also adopted the inclination, PA, RA, and DEC-offset values from Table \ref{tab:rotbestfit}.

We aligned the spectra across different azimuths to the same (stellar) peak velocity, accounting for the azimuthal dependence $\cos(\phi)$ of the Keplerian rotation in Eq. \ref{eq:1} using the Python package \texttt{eddy} \citep[][]{Teague(2019_eddy)}. Next, we averaged the aligned spectra to generate local spectral line profiles, thereby boosting the S/N of the local (i.e. per beam) line profile at a given disc radius, enabling an accurate analysis of its shape. However, one important caveat is that this method assumes the local spectral line shape at a given radius is independent of azimuth, which is not the case in our dataset (Fig. \ref{fig:12COmmaps} right panel). The largest variation occurs between $\sim$0$^{\circ}$ and $\pm$ 45$^{\circ}$, reaching a value of $\sim$300 m s$^{-1}$ at the peak gas surface density ($0\farcs61$-$0\farcs63$). Therefore, the S/N boost from this method comes at the cost of averaging out azimuthal variation in the line profile and smoothing the observed line profile.

The local line profiles for $^{12}\textrm{CO}$ (left panel of Fig. \ref{fig:stacked_spectra}) and $^{13}\textrm{CO}$ (right panel of Fig. \ref{fig:stacked_spectra}) are presented as a function of radial distance from the central star.\footnote{The local line profile error bars for both $^{12}\textrm{CO}$ and $^{13}\textrm{CO}$ have been re-scaled by a re-scaling value, which is fitted for in Section \ref{sec:1DResults}.} Both transitions exhibit broad local linewidths, with $^{12}\textrm{CO}$ displaying a full width half maximum (FWHM) of $\sim1092$ m s$^{-1}$ in the $0\farcs61$-$0\farcs63$ annulus ($\sim71$ - $74$ au), where its Gaussian shape is evident (Fig. \ref{fig:best_fit}). This broadening cannot be attributed to the spectral resolution of the $^{12}\textrm{CO}$ data, which, at 26 m s$^{-1}$, is ${\sim}42$ times smaller than the observed FWHM. The unexpected Gaussian shape and significant line broadening suggest the gas may be optically thin, contrary to the high optical depth expected from the literature \citep[e.g.][]{Moor(2017)}. In the optically thick case, we would in theory expect the profile to exhibit a top-hat shape \citep[e.g.][]{Horne(1986)}, as observed in optically thick protoplanetary discs \citep[e.g.][]{Erik(2018)}.

However, in protoplanetary discs, the local line profiles before azimuthal averaging can exhibit double-peaked emission due to the two-sided emitting layers commonly observed in these discs \citep{Izquierdo(2025)}. As a result, azimuthal averaging can make the overall line profile appear more Gaussian. In contrast, there is no evidence for two-sided emission layers in debris discs, where freeze-out and vertical temperature gradients are not expected. Furthermore, the $^{12}$CO channel maps show no indication of a two-sided emission structure at the spectrospatial resolution of our data. In Fig. \ref{fig:stacked_azimuth_river}, we show the line profiles as a function of azimuth, extracted from an annulus centred on the peak gas surface density ($0\farcs61$–$0\farcs63$). The profiles are approximately Gaussian in shape, even before azimuthal averaging.

\subsection{Local line profile modelling}
\label{sec:1DResults}

\begin{figure}
\centering
  \includegraphics[width=0.8\linewidth]{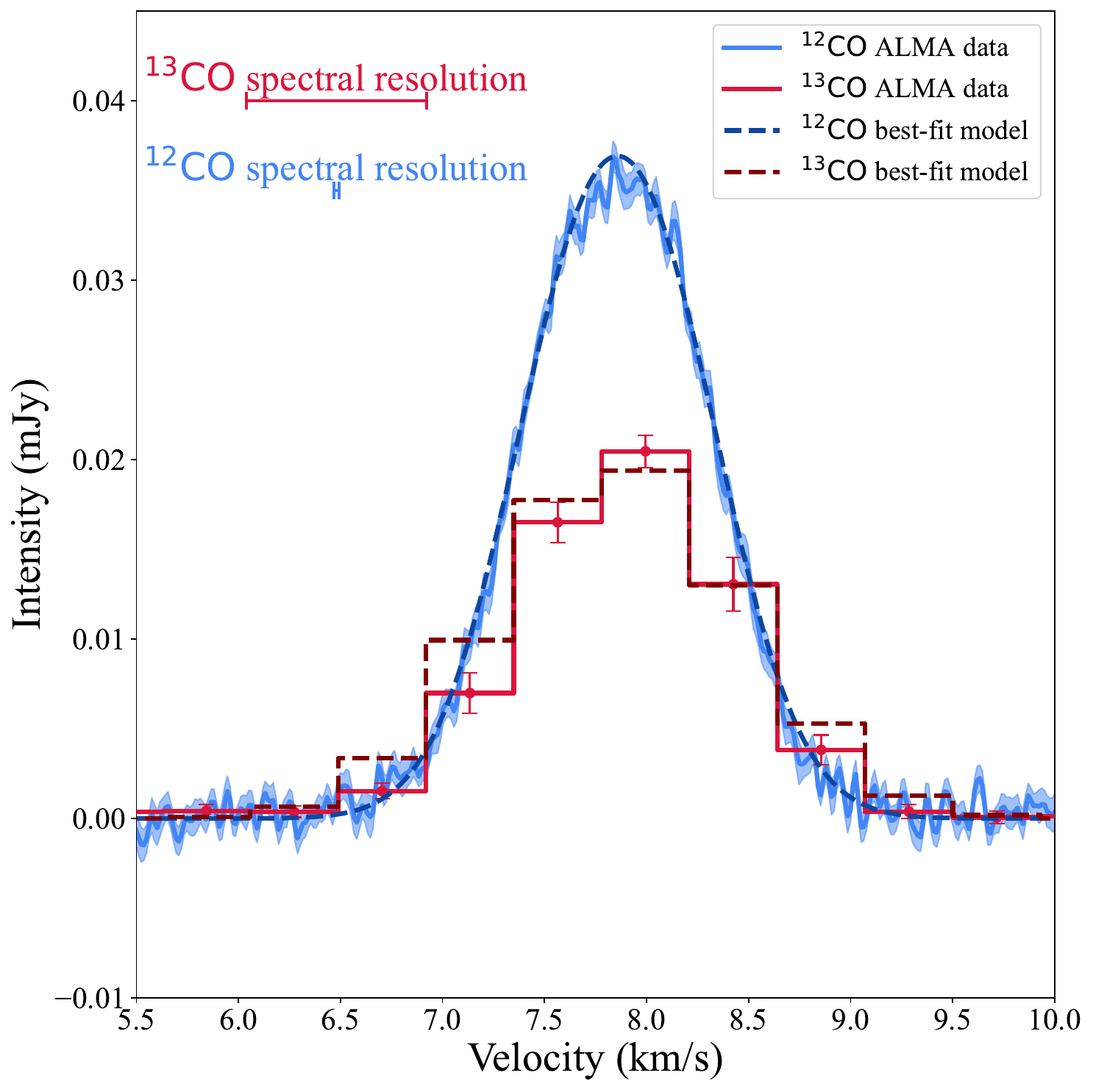}
  \caption{Local line profiles for $^{12}\textrm{CO}$ (blue) and $^{13}\textrm{CO}$ (red) created from an annulus extracted at the peak gas surface density ($0\farcs61$-$0\farcs63$). Best-fit models (dashed lines) are generated using best-fit values from Table \ref{table:best_fit_model}. Top left: Spectral resolution for $^{12}\textrm{CO}$ (blue) and $^{13}\textrm{CO}$ (red). Error bars for $^{12}\textrm{CO}$ (blue) and $^{13}\textrm{CO}$ (red) have been re-scaled by the fitted factor f$_{12}$ and f$_{13}$.}
  \label{fig:best_fit}
\end{figure}

To investigate the local CO gas properties and optical depth, we simultaneously fitted the local $^{12}\textrm{CO}$ and $^{13}\textrm{CO}$ line profiles with a simple radiative transfer model for each annulus. We modelled the local azimuthally averaged spectral line shape (e.g. Fig. \ref{fig:stacked_spectra} and \ref{fig:best_fit}) as arising from a homogeneous parcel of gas filling the beam in the absence of significant background radiation. In that case, the radiative transfer equation describing the (specific) intensity measured, assuming the dominant broadening mechanism is thermal Doppler broadening due to the gas kinetic temperature, reads

\begin{equation}
I_{\nu} = B_{\nu}(T_{\mathrm{exc}}) \left(1 - e^{-\tau(\nu)} \right),
\label{eq:inu}
\end{equation}

\noindent where $B_{\nu}(T_{\mathrm{exc}})$ is the blackbody intensity at the observed frequencies \( \nu \) and excitation temperature \( T_{\text{exc}} \), $\nu_0$ is the central frequency of the spectral line, and $\Delta \nu$ is the linewidth. The optical depth $\tau(\nu)$ is given by

\begin{equation}
\tau(\nu) = \tau_0  e^{\displaystyle -\frac{(\nu - \nu_0)^2}{\Delta \nu^2}},
\label{eq:tau}
\end{equation}

\noindent where $\tau_0$ is the optical depth at $\nu_0$ and the linewidth $\Delta \nu$ is given by $\Delta \nu = \frac{\Delta v}{c} \nu_0$, where $c$ is the speed of light and $\Delta v$ is given by

\begin{equation}
    \Delta v = \sqrt{\frac{2k_{B}T_{\mathrm{kin}}}m},
\end{equation}

\noindent where $k_{B}$ is the Boltzmann constant, $T_{\mathrm{kin}}$ is the kinetic temperature, and $m$ is the mass of a $^{12}\textrm{CO}$ or $^{13}\textrm{CO}$ molecule.

In our model, the emission is parametrised primarily by: the $^{12}\textrm{CO}$ optical depth, the $^{13}\textrm{CO}$ optical depth, the kinetic temperature (which sets the linewidth) and the excitation temperature (which determines the Planck function in Eq. \ref{eq:inu}). In our model, we fitted the excitation and kinetic temperature separately to approximately account for non-local thermodynamic equilibrium (non-LTE) effects. However, it is important to note that we calculate the level populations following the Boltzmann distribution, assuming a given excitation temperature, which we fitted for. We also assume that $^{12}\textrm{CO}$ and $^{13}\textrm{CO}$ have the same kinetic and excitation temperature. Additional free parameters include systemic velocity, which was used to shift the model spectrum, and two parameters, f$_{12}$ and f$_{13}$, which are constant values by which we multiplied the uncertainties to account for the inaccuracy of error bars in the local spectra of $^{12}\textrm{CO}$ and $^{13}\textrm{CO}$, respectively.

The emission models for both $^{12}\textrm{CO}$ and $^{13}\textrm{CO}$ were generated on the same velocity grid as the ARKS data. The generated models were then convolved with a Gaussian with a FWHM equal to the spectral resolution of the instrument (26 m s$^{-1}$ for $^{12}\textrm{CO}$ and 884 m s$^{-1}$ for $^{13}\textrm{CO}$). While ALMA’s spectral response is more complex than a Gaussian, this approximation allowed us to account for instrumental broadening in this simple model{\footnote{Information on ALMA's spectral response can be found at \url{https://safe.nrao.edu/wiki/pub/Main/ALMAWindowFunctions/Note_on_Spectral_Response.pdf}.}} Both $^{12}\textrm{CO}$ and $^{13}\textrm{CO}$ were simultaneously modelled across each annulus from $0\farcs32$ to $0\farcs98$, with each annulus having a width of $0\farcs02$. We fitted the emission line profiles using a Bayesian MCMC approach, as described in Section \ref{sec:ObservationsResults}.

Fig. \ref{fig:best_fit} shows the best-fit models for $^{12}\textrm{CO}$ (dashed blue line) and $^{13}\textrm{CO}$ (dashed red line), generated using the best-fit parameters from Table \ref{table:best_fit_model} for an annulus placed near the peak of the intensity radial profile at 73 au. The models are plotted alongside the data, showing that the $^{12}\textrm{CO}$ model provides a good fit. In contrast, the $^{13}\textrm{CO}$ model fails to reproduce the observed linewidth (Fig. \ref{fig:best_fit}), a consequence of assuming the same kinetic temperature for both isotopologues. This indicates that the pre-spectral-convolution linewidth of  $^{12}\textrm{CO}$ is broader than that of $^{13}\textrm{CO}$, which can be attributed to the fact that $^{13}\textrm{CO}$ has a much poorer spectral resolution. Furthermore, the excitation and kinetic temperatures are discrepant, with the kinetic temperature several hundreds of kelvin higher than the excitation temperature, indicating that the gas is in non-LTE under optically thin conditions.

Fig.~\ref{fig:best_params} shows how the best-fit width for the azimuthally averaged $^{12}\textrm{CO}$ profile varies as a function of radius. We only plot $^{12}\textrm{CO}$ here because the kinetic temperatures derived for both $^{12}\textrm{CO}$ and $^{13}\textrm{CO}$ are the same, and the small difference in mass results in very similar linewidths. We observe that the width for $^{12}\textrm{CO}$ is very large and decreases with distance. This decrease would make sense in theory if the linewidth is dominated by thermal Doppler broadening, since we observe a decrease in CO gas temperature with distance from the central star. However, attributing the decrease solely to temperature would require a steep radial dependence. Alternatively, it could be explained by Keplerian shear, which is also expected to decrease with radius (Appendix \ref{sec:keplerianshear}), or by a radial decrease in non-thermal broadening.
 
Although this toy model suggests that both $^{12}\textrm{CO}$ and $^{13}\textrm{CO}$ are optically thin, or marginally optically thin, at all radii, it does not account for Keplerian shear and fails to reproduce the observed $^{13}\textrm{CO}$ linewidth. Even though we find best-fit values for optical depths ${\lesssim} 1$ and high kinetic temperatures for both $^{12}\textrm{CO}$ and $^{13}\textrm{CO}$, the discrepancy between the model and the data suggests that our toy model is missing an important observational effect, such as the effect of Keplerian shear. 

\begin{table}
\centering
\caption{Best-fit values derived from fitting a simple toy model to an $^{12}\textrm{CO}$ and $^{13}\textrm{CO}$ local line profiles.}
\label{table:best_fit_model}
\begin{tabular}{l c}
\hline\hline
                              & Best-fit Value  \\ \hline
$^{12}\textrm{CO}$ $\tau_{0}$  & $0.50_{-0.33}^{+0.13}$ \\ 
$^{13}\textrm{CO}$ $\tau_{0}$  & $0.30_{-0.32}^{+0.12}$ \\ 
T$_{\textrm{exc}}$ (K)         & $80_{-14}^{+66}$    \\ 
T$_{\textrm{kin}}$ (K)         & $600_{-27}^{+38}$  \\ 
v$_{\textrm{BARY}}$ (m s$^{-1}$) & $7858_{-3}^{+2}$   \\ 
ln (f$_{12}$)                 & $-1.4_{-0.12}^{+0.02}$ \\ 
ln (f$_{13}$)                 & $-1.1_{-0.13}^{+0.12}$ \\ 
\hline
\end{tabular}
\tablefoot{The local line profiles were created from an annulus extracted at the peak gas surface density ($0\farcs61$-$0\farcs63$). The quoted best-fit values are the median of the marginalised distribution. The uncertainties are based on the 16th and 84th percentiles of the marginalised distributions.}
\end{table}

\section{3D radiative transfer modelling}
\label{sec:RADMC-3D}

Since the simple toy model described in Section \ref{sec:1DResults} does not account for the mixing of velocities within the beam or geometric effects along the line of sight in each beam, we created a full radiative transfer model of the CO emission using \texttt{RADMC-3D} \citep[]{RADMC3D(2012)}. This model incorporates these effects to either confirm or challenge the results from Section \ref{sec:1DResults}. Our model takes an assumed 3D CO density, temperature, and velocity structure and solves the radiative transfer equation to produce a simulated CO line emission cube. In Section \ref{sec:Optically Thin Model}, we describe an optically thin $^{12}$CO \texttt{RADMC-3D} model created to test whether we can reproduce both the results from Section \ref{sec:1DResults} and the observed data. Our aim is not to perform a detailed fit of the $^{12}$CO cube, which would be computationally intensive due to the large number of spectral channels, but rather to verify whether the results from our simpler model are consistent with one that includes 3D geometry and beam convolution effects. Then, in Section \ref{sec:13CO RADMC3D Modell} we describe a different approach in which we directly fitted an \texttt{RADMC-3D} model to the $^{13}$CO data cube. In Section \ref{sec:12CO RADMC3D Modell}, we then test whether the best-fit parameters derived for $^{13}$CO provide a good fit to the $^{12}$CO by simply re-scaling the model mass. 

\subsection{Optically thin model}
\label{sec:Optically Thin Model}

\subsubsection{Model assumptions}

For the 3D density structure, we modelled the CO gas surface density as a Gaussian with $r_{\rm peak} = 73$ au and $\sigma=21$ au, where both parameters are fixed. These values were determined by assuming the gas is optically thin, where $I_{\nu} \propto e^{\frac{-(r - r_\mathrm{peak})^2}{2\sigma^2}} r^{-0.5}$. Using this equation, we fitted the radial profile (Fig. \ref{fig:radial_profile}) and determined the standard deviation $\sigma$ and peak radius $r_{\rm peak}$. The surface density distribution was normalised such that its 2D integral yields the total CO mass, treated as a free parameter. 

We assumed a Gaussian vertical profile with a vertically constant temperature, as expected from hydrostatic equilibrium. For a disc in hydrostatic equilibrium, at a given radius, the scale height reads

\begin{equation}
    H = \sqrt{\frac{k_{b} T(r) r^{3}}{\mu m_{p} G  M_{\ast }}},
    \label{eq:3}
\end{equation}

where $T$ is the temperature distribution with radius, $m_{p}$ is proton mass. In Eq. \ref{eq:3}, the vertical scale height was determined by the temperature and the mean molecular weight $\mu$, which we adjusted to match the observed data (Section \ref{sec:RADMC-3D Modelling}). For simplicity, we assumed local thermodynamic equilibrium (LTE) to calculate the CO level populations. For the velocity distribution, we assumed Keplerian rotation around a star with a mass of 1.89 M$_{\ast}$ (Table \ref{tab:rotbestfit}). The radiative transfer was then run using the inclination, PA, RA, and DEC-offset values from Table \ref{tab:rotbestfit}, and a distance to the star of 117.9 pc \citep[][]{Gaia(2023)} to produce our $^{12}\textrm{CO}$ model cube. 

We created a model cube matching the spectrospatial coordinates of the data, with 678 $\times$ 678 $0\farcs0225$ pixels with 1182 13~m~s$^{-1}$ channels. For direct comparison with the data, we spectrally convolved the model with a Gaussian (FWHM = 2 $\times$ channel width) to account for ALMA's native spectral resolution. Finally, to simulate the spatial resolution of the data, the model cube was spatially convolved with a 2D Gaussian kernel parametrised by the major and minor axes of the observational beam ($0\farcs12$~$\times$~$0\farcs11$).

\subsubsection{Data comparison}
\label{sec:RADMC-3D Modelling}

\begin{figure}
\centering
  \includegraphics[width=1\linewidth]{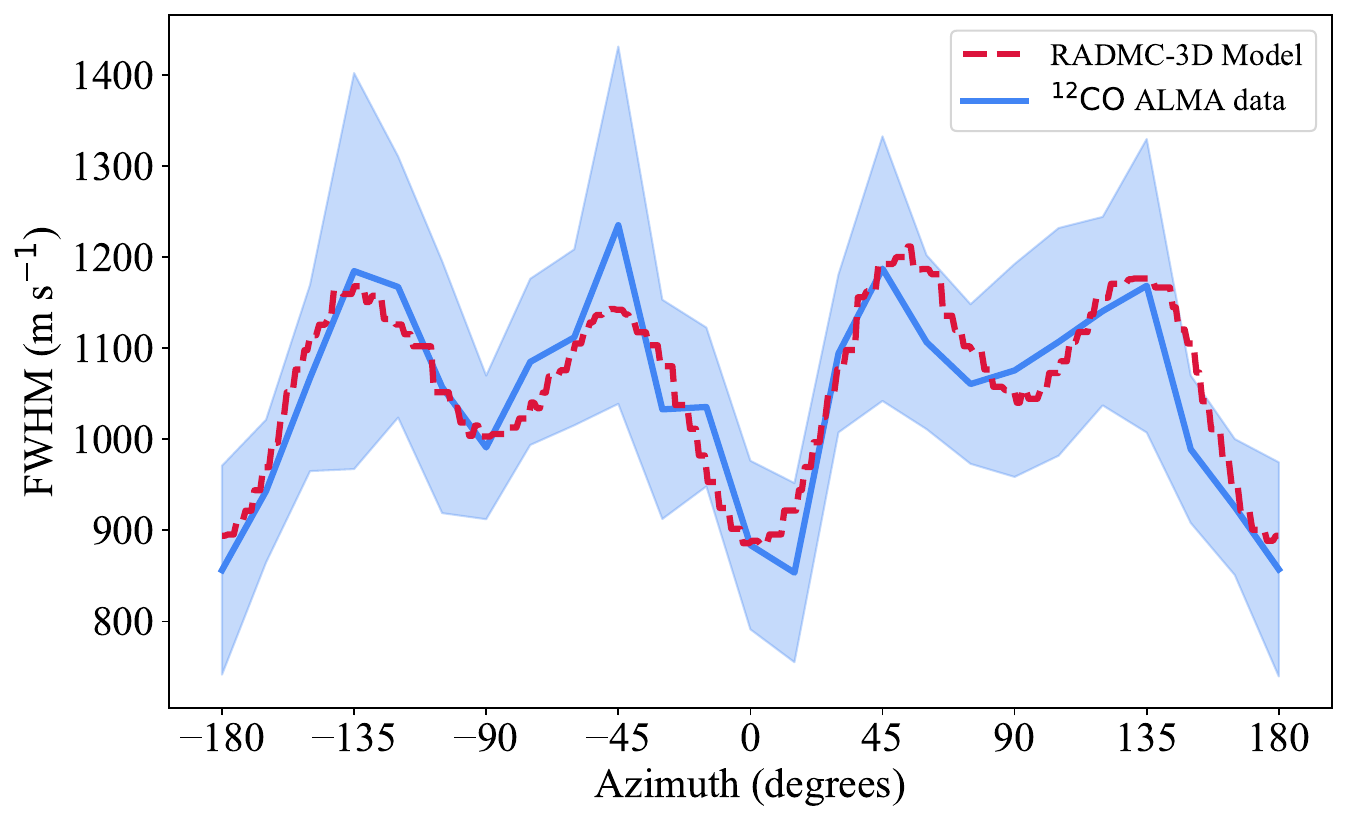}
  \caption{Comparison of FWHM (m s$^{-1}$) for $^{12}\textrm{CO}$ ALMA data (blue line) and the \texttt{RADMC-3D} optically thin model (dashed red line), where the FWHM is derived by fitting a Gaussian to each spectrum (assuming azimuthal sections of 15$^\circ$ for the data) using an MCMC approach for both the model and data. For both the model and the data, an annulus between $0\farcs61$ and  $0\farcs63$ (corresponding to the disc’s peak radial surface density) is extracted. The error bars for the $^{12}\textrm{CO}$ ALMA data (shaded regions) have been re-scaled by a constant factor, which was fitted for.}
  \label{fig:FWHM}
\end{figure}

\begin{figure}
\centering
  \includegraphics[width=0.9\linewidth]{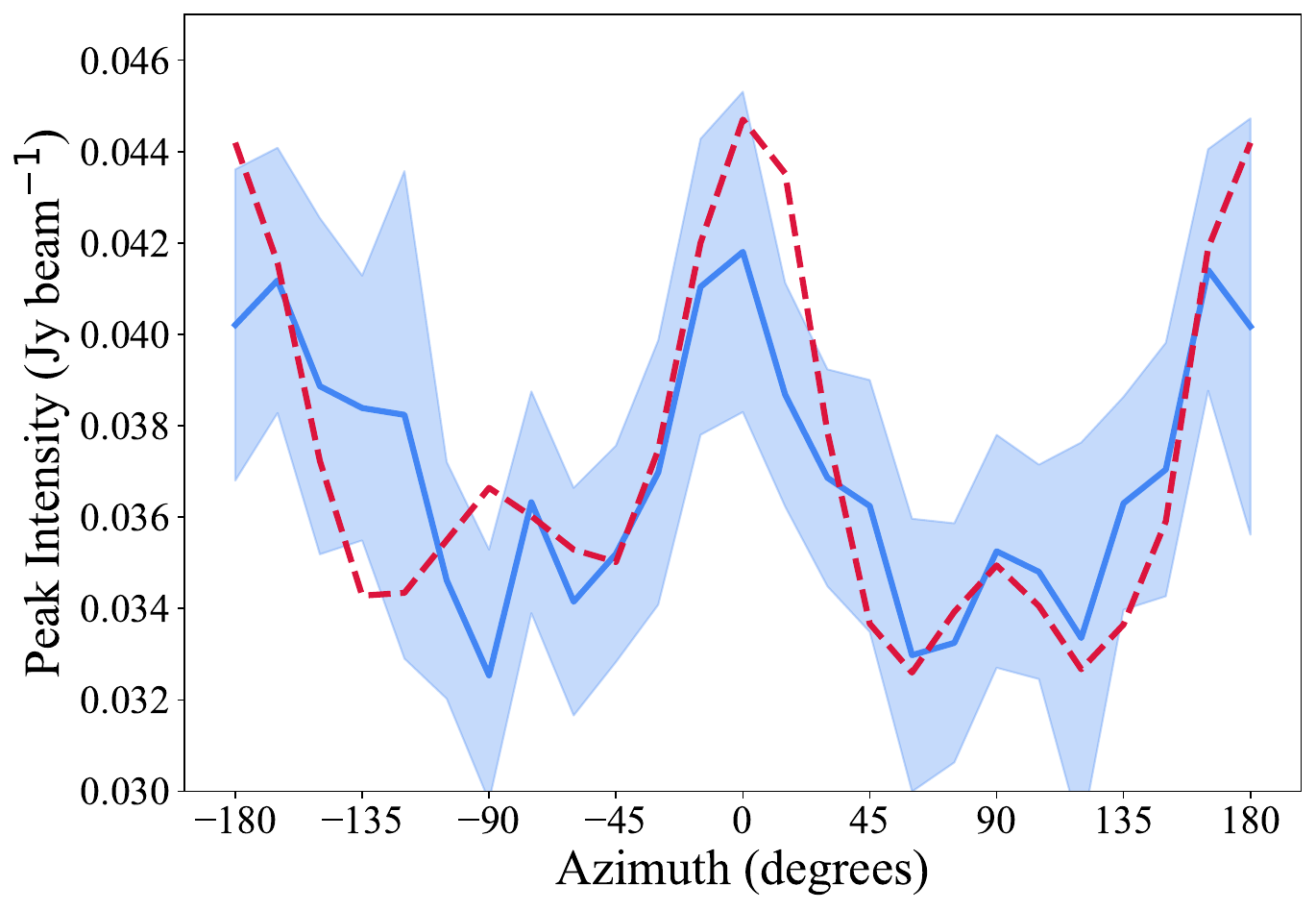}
  \caption{Peak intensity (Jy beam$^{-1}$) for the $^{12}\textrm{CO}$ data (blue line) and the \texttt{RADMC-3D} optically thin model (dashed red line), with the peak intensity for each azimuth (assuming azimuthal sections of 15$^\circ$ for the data) derived by fitting a Gaussian to each spectrum across all azimuths using an MCMC approach for both the model and data. For both the model and the data, an annulus between $0\farcs61$ and  $0\farcs63$  (corresponding to the disc’s peak radial surface density) is extracted. The error bars for the $^{12}\textrm{CO}$ ALMA data (shaded regions) have been re-scaled by a constant factor, which was fitted for.}
  \label{fig:Peak_Brightness}
\end{figure}

\begin{figure}
\centering
  \includegraphics[width=0.8\linewidth]{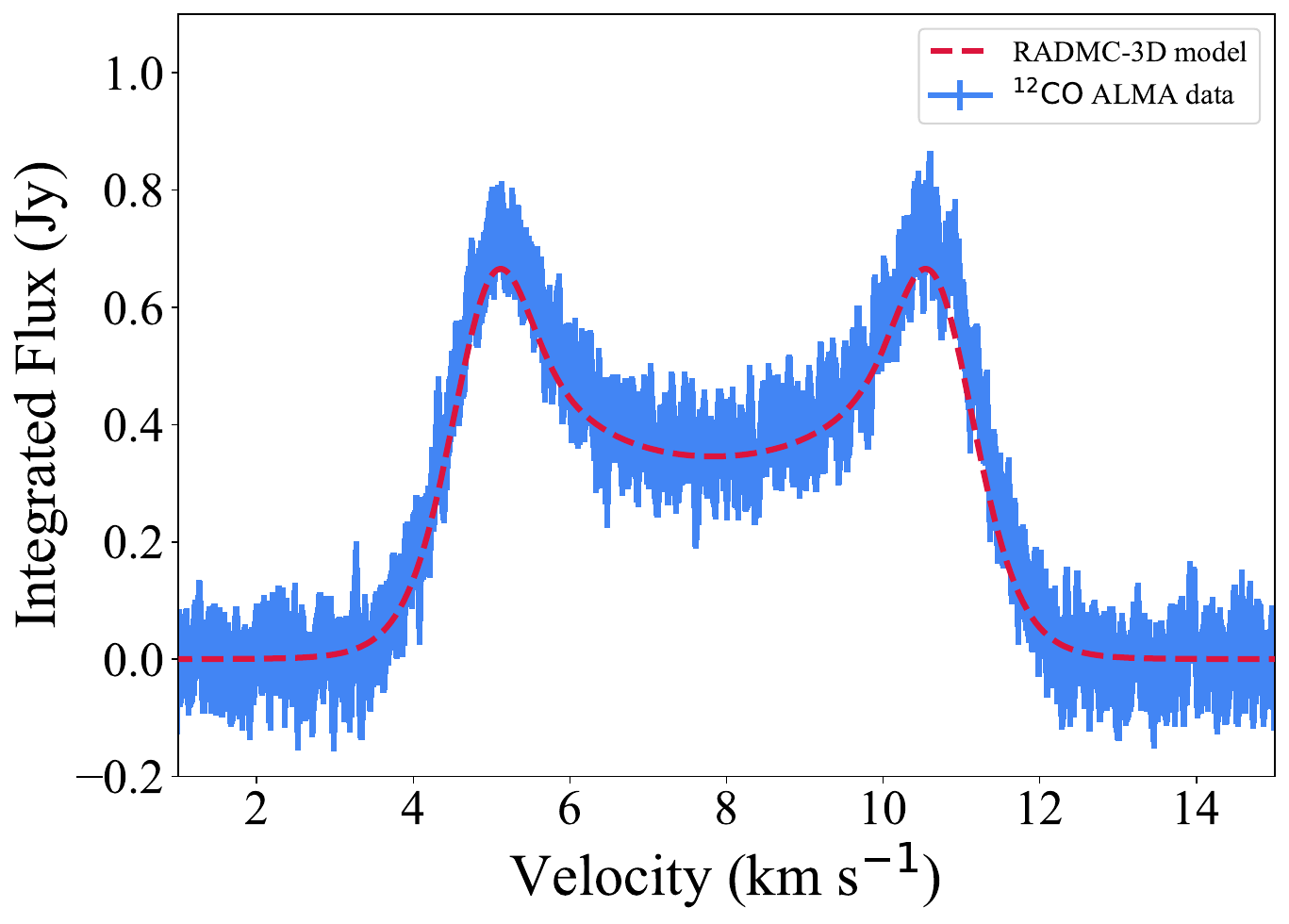}
  \caption{1D spectra for $^{12}\textrm{CO}$ (blue line) and \texttt{RADMC-3D} optically thin model (dashed red line) showing the integrated flux as a function of velocity.}
  \label{fig:1D}
\end{figure}

We varied three parameters in our \texttt{RADMC-3D} model: $\mu$, the mean molecular weight; $T_{\rm peak}$ (K), the temperature at the peak radial surface density (under the assumption of LTE, the excitation and kinetic temperatures are equivalent); and $M_{\rm CO}$ (M$_{\oplus}$), the CO mass. Since we did not formally fit the cube, these parameters were adjusted to achieve the best qualitative agreement with key features of our dataset. This includes the spectral width (FWHM) of the line (specific) intensity and its azimuthal dependence (Fig. \ref{fig:FWHM}), the peak intensity and its azimuthal dependence (Fig. \ref{fig:Peak_Brightness}), and the 1D spectral line profile obtained by spatially integrating across the entire disc between a radius of 0 and 200 au (Fig. \ref{fig:1D}). While the spectral width and peak intensity allow us to explore the linewidth of $^{12}\textrm{CO}$, the potential effects of geometry, and the effect of Keplerian shear, the 1D spectral line profile of the whole disc in Fig. \ref{fig:1D} ensures that the model’s CO mass, as well as its radial and velocity structure remain representative of the entire disc.

We find that an optically thin model with a peak optical depth of $\sim$0.11, $T_{\rm peak} = 380$ K, $\mu = 28$ (CO dominated), and $M_{\rm CO} = 10^{-3}$ M$_{\oplus}$ reproduces many key features of the data. Figures \ref{fig:FWHM}, \ref{fig:Peak_Brightness}, and \ref{fig:1D} compare the best qualitative \texttt{RADMC-3D} model (dashed red line) with the $^{12}\textrm{CO}$ data (blue line). For both the model and the data, an annulus between  $0\farcs61$ and  $0\farcs63$ (corresponding to the disc's peak radial surface density) is extracted for analysis, as detailed in Section \ref{sec:Local Line Profile Extraction}. We fitted a Gaussian with FWHM and peak intensity parameters to the spectra at different azimuths (assuming azimuthal sections of 15$^\circ$ for the data) with an MCMC approach, as described in Section \ref{sec:1DResults}. We also fitted an additional parameter for the data multiplied by uncertainties that are inaccurate when compared to the RMS scatter in noise-only regions of the local line profiles. The fitting procedure consists of 2000 steps with a burn-in period of 100 steps due to faster convergence.

In Fig. \ref{fig:FWHM}, we see that the \texttt{RADMC-3D} model reproduces the FWHM derived from the $^{12}\textrm{CO}$ data well. The derived temperature of 380 K is notably lower than the value of 600 K determined in Section \ref{sec:1DResults} but still significantly higher than the expected blackbody temperature of $\sim$63 K at 73 au, given a stellar luminosity of 14 L$_{\odot}$ \citep{Matra(2025)}, assuming the dust and gas have the same temperature. Additionally, we observe azimuthal variation in the $^{12}\textrm{CO}$ linewidth (Fig. \ref{fig:FWHM}), which shows two deeper minima along the semi-major axis, two shallower minima along the semi-minor axis, and maxima at $\pm45^{\circ}$ and $\pm135^{\circ}$. This pattern corresponds to the X-shaped feature in the linewidth map (Fig. \ref{fig:12COmmaps}, right panel). The variation arises due to Keplerian shear, which creates a projected velocity gradient both radially and azimuthally across a single beam. The degree of shear depends on the azimuthal position at a given disc radius, as shown analytically in Appendix \ref{sec:keplerianshear} and captured by our \texttt{RADMC-3D} model. The combination of radial and azimuthal shear produces the observed FWHM variation (Fig. \ref{fig:analyticlocalprofiles}), demonstrating that spectral line broadening due to the effect of Keplerian shear is significant at the disc's surface density peak and depends on the azimuth. However, while Keplerian shear explains the azimuthal variation, its broadening effect is not strong enough to avoid the need for a high temperature.

Both temperature and $\mu$ also influence the azimuthal dependence of the FWHM (and line peak intensity) through the disc's vertical scale height. In Eq. \ref{eq:3}, an increase in $T$ or a decrease in $\mu$ (i.e. the gas is dominated by lighter molecules, such as H$_{2}$ instead of CO) increases the scale height, making the disc more vertically extended. In a vertically extended disc, the line of sight intersects a broader range of velocities at different heights and radii due to the disc’s non-zero inclination. As a result, a greater range of velocities is probed along the line of sight, enhancing the effect of Keplerian shear and leading to larger FWHM across all azimuths. 

Fig. \ref{fig:Peak_Brightness} shows the change in peak intensity with azimuth for the \texttt{RADMC-3D} model (dashed red line) and data (blue line). We see that our \texttt{RADMC-3D} model reproduces the azimuthal variation of the peak intensity in the data. Increasing the input mass (and thus the column density within a beam) or decreasing the temperature causes the spectral line peaks within a beam to rise, which in turn increases the peak intensity; thus, Fig. \ref{fig:Peak_Brightness} is jointly constrained by mass and temperature. The integrated line flux of the model also closely matches that of the data (Fig. \ref{fig:1D}), which constrains the mass of $^{12}\textrm{CO}$ in the system to $1\times10^{-3}$ M$_{\oplus}$, with the temperature separately constrained by the linewidth.

In Fig. \ref{fig:Peak_Brightness}, we note that the model exhibits local azimuthal peaks at $\pm 90^{\circ}$, also seen in the peak intensity map for the optically thin model presented in this section (Fig. \ref{fig:opticalthin_mom8}). These local azimuthal peaks at $\pm 90^{\circ}$ are not observed in the peak intensity map of the data (Fig. \ref{fig:12COmmaps}, centre). Despite the uncertainties of the single 73 au annulus analysed here and the peak intensity map, these local azimuthal peaks at $\pm 90^{\circ}$, if present, should be at least marginally resolved in the peak intensity map. As a result, any optically thin model such as ours should not produce an X-shaped pattern in the velocity integrated intensity map (see Fig. \ref{fig:opticalthin_mom0}), which is in strong disagreement with our observed data (Fig. \ref{fig:12COmmaps}, left). One potential explanation for this discrepancy is the assumption of LTE in the \texttt{RADMC-3D} model, which couples the linewidth and peak intensity (through the J=3 upper-level population). If the system is outside of LTE, as suggested in Section \ref{sec:Local Spectral Line Profiles}, the level population and peak intensity would decouple from the linewidth, potentially allowing for a better fit than the model herein permits. In summary, although the optically thin model can reproduce many features of the data, it requires unrealistically high temperatures, cannot account for the X-shape, and, as demonstrated in Section \ref{sec:Local Spectral Line Profiles}, overestimates the $^{13}$CO linewidths.

\subsection{\texttt{RADMC-3D} modelling}

\subsubsection{$^{13}\textrm{CO}$ \texttt{RADMC-3D} model}
\label{sec:13CO RADMC3D Modell}

While a strength of the optically thin model is that it can immediately explain the azimuthally averaged Gaussian line shapes observed (Section \ref{sec:Local Spectral Line Profiles}), its key shortcomings are the inability to reproduce the X-shape in the velocity integrated intensity map and the smaller linewidth of $^{13}\textrm{CO}$ compared to $^{12}\textrm{CO}$ (after accounting for spectral resolution). These are features expected for optically thick lines, for example, the X-shape in the velocity integrated intensity map is expected if the peak intensity map is saturated by optical depth, after combination with the X-shaped linewidth map (caused by Keplerian shear). Additionally, high optical depth effects naturally produce broader linewidths for more optically thick species, potentially explaining the broader $^{12}\textrm{CO}$ compared to $^{13}\textrm{CO}$. However, the challenge for an optically thick model would be the Gaussian local line profiles, where the local line profiles are Gaussian at each azimuth before azimuthal averaging.

For simplicity, we assumed LTE to calculate the CO level populations. For the velocity distribution, we assumed Keplerian rotation around a star with a mass of 1.89 M$_{\ast}$ (Table \ref{tab:rotbestfit}). The radiative transfer was then run using the inclination, PA, RA, and DEC-offset values from Table \ref{tab:rotbestfit} and a distance to the star of 117.9 pc \citep[][]{Gaia(2023)} to produce our $^{12}\textrm{CO}$ model cube. 

We fitted a \texttt{RADMC-3D} model to the $^{13}\textrm{CO}$ data cube, leaving the following free parameters: $R$ (au), the peak of the Gaussian surface density distribution; $\Delta R$ (au), the FWHM of the Gaussian distribution; $T_{73}$ (K), the temperature at 73~au; $\beta$, the temperature power-law index; $\mu$, the mean molecular weight; $M_{\rm CO}$ (M$_{\oplus}$), the CO mass; $i$ (deg), the inclination; $\mathrm{PA}$ (deg), the position angle; and M$_{\star}$ (M$_{\odot}$), the stellar mass. Following Section \ref{sec:Optically Thin Model}, we accounted for ALMA's native spectral resolution and simulated the spatial resolution of the data. The fit to the $^{13}$CO cube was carried out using MCMC, with an affine-invariant ensemble sampler implemented in the emcee package (Goodman \& Weare 2010; Foreman-Mackey et al. 2013), with the uncertainty per pixel and channel assumed to be equal to the cube's RMS multiplied by the square root of the number of pixels per beam and the number of channels per effective bandwidth to account for spectrospatial correlation \citep[e.g.][]{Marino(2016)}.

\begin{figure*}
\centering
  \includegraphics[width=.95\linewidth]{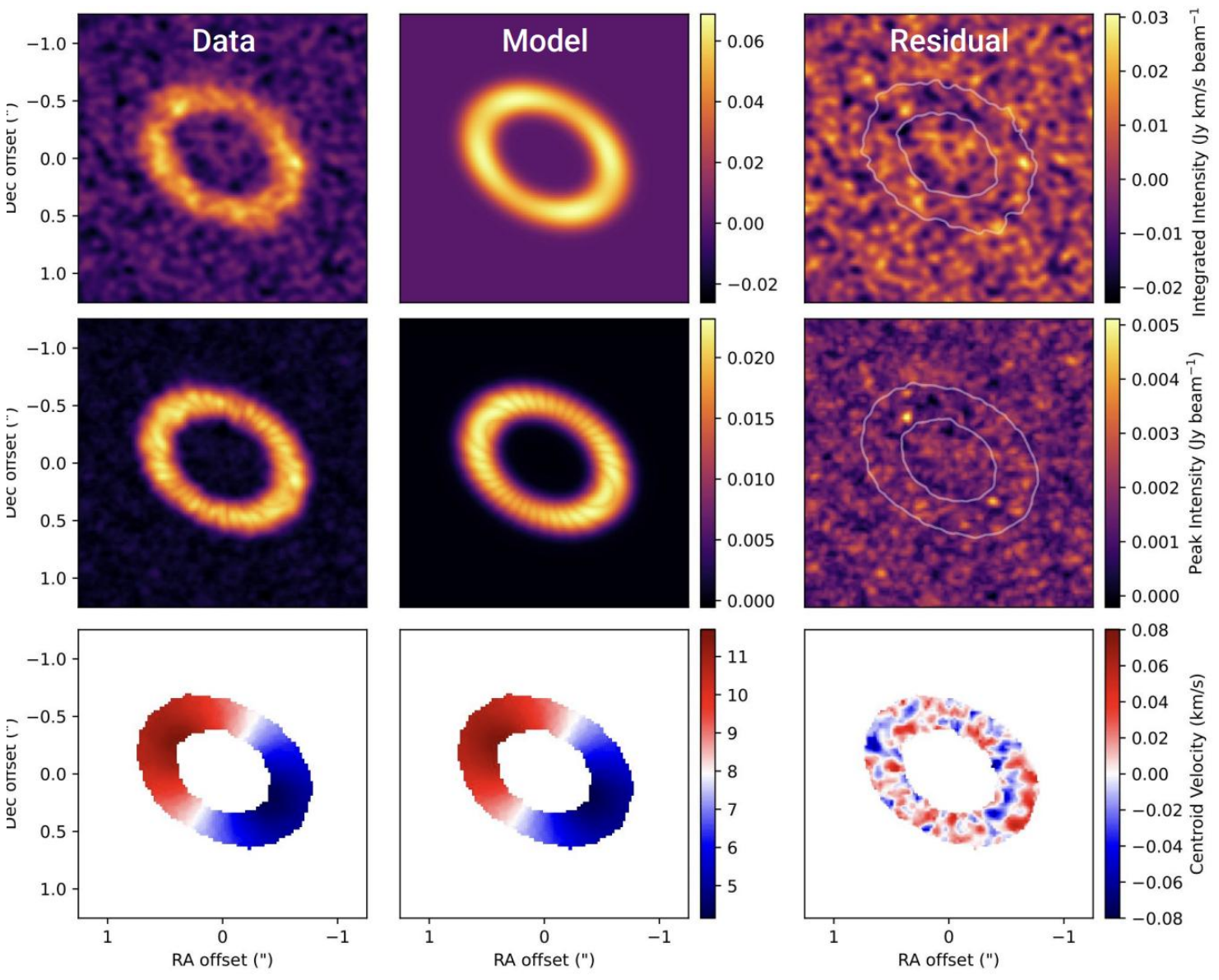}
  \caption{Left: $^{13}$CO data. Middle: Best-fit \texttt{RADMC-3D} optically thick model. Right: Residuals (data minus model). From top to bottom: Velocity integrated intensity, peak intensity, and velocity maps, respectively.}
  \label{fig:HD121617_13CO_gauss_bestfitmoments}
\end{figure*}

In Fig.~\ref{fig:HD121617_13CO_gauss_bestfitmoments}, we present the best-fit \texttt{RADMC-3D} model, generated using the parameters listed in Table \ref{table:best_fit_geometry_13CO}, using the corner plots shown in Fig. \ref{fig:13CO_gauss_corner}. The left column displays the $^{13}$CO data, the middle column shows the best-fit model, and the right column presents the residuals (data minus model). The top row shows the velocity integrated intensity map, the middle row shows the peak intensity map, and the bottom row is the velocity map. The best-fit model with a temperature of 38~K and a $^{13}$CO mass of $2 \times 10^{-3}$ M$_{\oplus}$ is a very good fit to the data (see also channel map residuals in Fig.\ref{fig:13COchannelmap}).

The best-fit model is optically thick, with a peak optical depth of 26, and as expected, successfully reproduces the X-shaped structure in the velocity integrated intensity map, which the optically thin model could not. We also find that, for an optically thick model to reproduce the data, the CO radial distribution must be narrower (FWHM of 17.3~au) than that of the observed intensity radial profile (48~au). This makes the $^{13}$CO radial width comparable to that of the dust \citep[14.1$\pm 1.2$~au,][]{Yinuo_arks}.

\begin{table}
\centering
\caption{Best-fit geometric and physical parameters derived from fitting a \texttt{RADMC-3D} model to the $^{13}$CO cube.}
\label{table:best_fit_geometry_13CO}
\begin{tabular}{l c}
\hline\hline
Parameter & Best-fit Value \\ \hline
$R$ (au) & $74.4_{-0.2}^{+0.2}$ \\
$\Delta R$ (au) & $17.3_{-0.9}^{+0.9}$ \\
$T_{73\ \mathrm{au}}$ (K) & $38.1_{-1.2}^{+1.3}$ \\
$\beta$ & $-0.14_{-0.07}^{+0.07}$ \\
$\mu$ & $12.6_{-1.1}^{+1.3}$ \\
$\log M_{^{13}\mathrm{CO}}$ (M$_\oplus$) & $-2.7_{-0.1}^{+0.2}$ \\
$i$ ($^\circ$) & $44.4_{-0.2}^{+0.2}$ \\
PA ($^\circ$) & $59.67_{-0.08}^{+0.08}$ \\
$M_\star$ (M$_\odot$) & $1.915_{-0.009}^{+0.009}$ \\
\hline
\end{tabular}
\tablefoot{Quoted values represent the medians of the marginalised posterior distributions. Uncertainties correspond to the 16th and 84th percentiles.}
\end{table}

In an optically thick model, the mass and kinetic temperature are degenerate as both cause an increase in linewidth. However, assuming LTE allows us to break this degeneracy by assuming that the kinetic temperature is the same as the excitation temperature (38\,K), which for an optically thick model is constrained by the peak intensity. In turn, this implies that the $^{13}$CO mass derived is strictly based on our assumption of LTE and on our choice of model.

Additionally, we fitted for $\mu$, the mean molecular weight, and derived a best-fit value of $12.6_{-1.1}^{+1.3}$. In our model, $\mu$ sets the vertical width of the gas distribution for a given kinetic temperature. Given that the model is optically thick, the observed intensity traces an elevated, vertically thin layer corresponding to $\tau \sim1$. The height of this layer at a given radius in our model depends on $\mu$, and on the CO column density and (vertically constant) temperature at that radius \citep[e.g.][]{Rosotti(2025)}. As the model linewidths and intensities separately constrain the latter two quantities, the constraint on $\mu$ likely comes from measuring a small vertical displacement of the $^{13}$CO emitting layer from the midplane. While this elevated layer is clearly seen in channel maps for protoplanetary discs \citep[e.g.][]{Rosenfeld(2013)}, we see no clear evidence for it in our $^{12}$CO or $^{13}$CO velocity maps or channel maps (Figs. \ref{fig:12COmmaps}, \ref{fig:13COmmaps}, \ref{fig:12COchannelmap}, \ref{fig:13COchannelmap}). Therefore, although our derived $\mu$ value is in agreement with the result of \citet{Seba_2arks} on the same dataset and is suggestive of a secondary origin for the gas, we caution that its derivation is critically dependent on our LTE assumption, as well as on other model choices such as the employed radial and vertical gas and temperature distribution.

\subsubsection{$^{12}\textrm{CO}$ \texttt{RADMC-3D} model}
\label{sec:12CO RADMC3D Modell}

\begin{figure*}
\centering
  \includegraphics[width=0.8\linewidth]{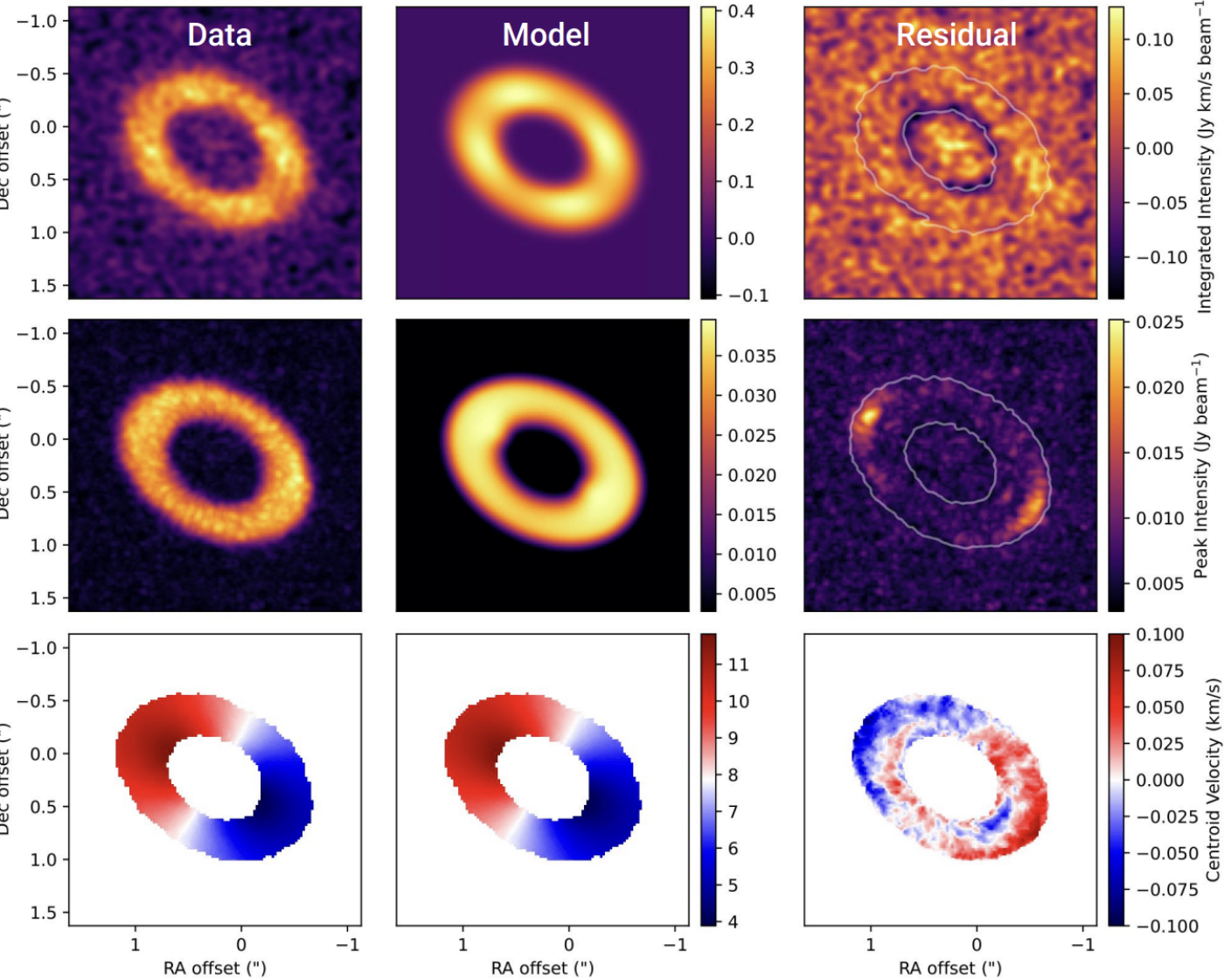}
  \caption{Left: $^{12}$CO data. Middle: Re-scaled optically thick model. Right: Residuals (data minus model). From top to bottom: Velocity integrated intensity, peak intensity, and velocity maps, respectively.}
  \label{fig:HD121617_12CO_gauss_bestfitmoments}
\end{figure*}

To investigate whether the $^{12}$CO data can also be reproduced by the same $^{13}\textrm{CO}$ optically thick model, we created a $^{12}\textrm{CO}$ \texttt{RADMC-3D} model using the $^{13}\textrm{CO}$ best-fit parameters summarised in Table~\ref{table:best_fit_geometry_13CO}. We scaled the mass by the ISM abundance ratio of 77, leading to a peak optical depth for the $^{12}$CO model of 3658. We find that, although some residuals remain in the maps (Fig. \ref{fig:HD121617_12CO_gauss_bestfitmoments}), the model reproduces the data reasonably well given that the mass was simply scaled. In the velocity integrated intensity map residuals, the model slightly over-predicts the intensity in the inner disc. This is due to the imperfect assumption of a perfectly symmetric Gaussian CO distribution. In the velocity maps residuals we observe structure, which we attribute to deviations from purely Keplerian rotation. In \citet{Seba_2arks}, a Keplerian model was fitted to the $^{12}\textrm{CO}$ velocity map. They found the same residual structure seen in Fig.~\ref{fig:HD121617_12CO_gauss_bestfitmoments}. In \citet{Seba_2arks}, this non-Keplerian behaviour was attributed to the density and pressure profile of a narrow gas ring. In the peak intensity map residuals, the model under-predicts the peak intensity at the ansae. This is due to the non-Keplerian motion of the gas, where discrepancies between the model's Keplerian velocity and the true gas velocity cause the model spectra in each pixel to be offset from the data. As a result, when the model is subtracted, residual peaks remain. Therefore, we conclude that our $^{13}\textrm{CO}$ best-fit model can reproduce the bulk of the $^{12}\textrm{CO}$ emission if simply re-scaled by ISM abundance ratios, although it is missing necessary details such as non-Keplerian rotation. We, however, caution that different total CO masses/abundance ratios may also fit the data, given the high optical depth of both isotopologues.

\section{Discussion}
\label{sec:Discussion}

\subsection{CO mass}
\label{sec: Low CO mass}

In CO-rich debris discs such as HD~121617, total CO mass estimates are typically derived from observations of rarer isotopologues, assuming ISM-like ratios. For instance, the CO mass derived for HD~121617 is $1.8 \pm 0.3 \times 10^{-2} $ M$_{\oplus}$ and $2.0 \pm 0.7\times 10^{-2}$ M$_{\oplus}$, based on $^{13}$CO and C$^{18}$O line fluxes, respectively \citep[][]{Moor(2017)}. Additionally, \citet{Cataldi(2023)} reports a CO mass of $2.57 \pm 0.04 \times 10^{-2}$ M$_{\oplus}$ using a simple LTE disc model fitted to disc-integrated fluxes. In comparison, our spectrospatially resolved ARKS analysis yields a best-fit $^{13}$CO mass of $2 \times 10^{-3}$ M$_{\oplus}$, which corresponds to a total CO mass of $1.48 \times 10^{-1}$ M$_{\oplus}$ assuming an interstellar $^{12}$CO/$^{13}$CO abundance ratio of 77. This value is higher than previous estimates in the literature. The discrepancy may arise from differences in model assumptions or due to C$^{18}$O also being moderately optically thick (as our $^{13}$CO model would suggest, assuming ISM $^{13}\textrm{CO}$/C$^{18}$O ratios apply). It is important to note that both $^{13}$CO and $^{12}$CO emission are optically thick, leading to degeneracies in 3D structure, temperature and velocity distribution, mass, and isotopologue abundance ratios, which are difficult to break without assumptions, even with high spectrospatial resolution observations.

To investigate the implications for the derived CO mass, we first calculated the peak $^{12}\mathrm{CO}$ surface density in our 3D radiative transfer model, which gives a vertical column density of $\sim$ $2\times 10^{19}$ cm$^{-2}$ at the radial peak of the disc. We then calculated the level of self-shielding\footnote{We used the shielding functions from the Leiden database found at \url{https://home.strw.leidenuniv.nl/~ewine/photo/references.html}} from photodissociation from the ISRF experienced by a CO molecule in the midplane along the vertical direction \citep{Visser(2009)}. Assuming CO is self-shielding only (i.e. absence of other shielding species), we find a CO lifetime of $\sim$0.07 Myr. Assuming a primordial gas origin where large amounts of unseen H$_{2}$ shield CO from photodissociation, the CO lifetime would become larger than the lifetime of the star, assuming a CO/H$_{2}$ ratio of $1\times10^{-4}$ \citep[][]{Kevin_Smith(2025)}. Therefore, the CO mass derived here for HD~121617 allows for a primordial gas origin in the presence of large amounts of H$_{2}$, which could, however, be at odds with the high mean molecular weights for the gas derived here (Section \ref{sec:13CO RADMC3D Modell}).

\subsection{$^{12}\mathrm{CO}/^{13}\mathrm{CO}$ line ratio and broad local line profiles}
\label{sec: Line Ratio}

As we find that both $^{12}$CO and $^{13}$CO are optically thick, the observed $^{12}\textrm{CO}/^{13}\textrm{CO}$ integrated line intensity ratio of $\sim$2.5 at the peak surface density does not directly trace the abundance ratio, which is instead largely consistent with ISM values. This intensity ratio changes little with radius, as reported by \citet{Sorcha_arks}, who also find similarly low $^{12}\textrm{CO}/^{13}\textrm{CO}$ line ratios in four other CO-rich exo-Kuiper belts (HD~9672, HD~32297, HD~131488, and HD~131835). Our observed $^{12}\textrm{CO}/^{13}\textrm{CO}$ line ratio of $\sim$2.5 is comparable to those obtained for gas-rich protoplanetary discs observed in the ALMA MAPS survey \citep[][]{Zhang(2021a)}, where both $^{12}$CO and $^{13}$CO are known to be optically thick. Therefore, our results for HD~121617 support a scenario where the other four CO-rich exo-Kuiper belts are also optically thick in $^{12}$CO and $^{13}$CO. Moreover, the X-shape seen in several of their velocity-integrated intensity maps corroborates that both lines are optically thick.

In our \texttt{RADMC-3D} model, we accounted for Keplerian shear, projection, and optical depth effects, and ALMA’s spectrospatial response and found that Keplerian shear and high optical depths cause the observed, surprisingly broad, linewidths. Additionally, we find that perfectly Gaussian line profiles in azimuthally averaged spectra can be obtained even if the line departs from Gaussian locally, because of the azimuthal dependence of width caused by Keplerian shear. Therefore, we caution against using the line profile shape to distinguish between optically thin or thick emission. We also conclude that accessing thermal broadening due to kinetic temperature, as well as non-thermal broadening, is challenging when using optically thick emission tracers, even at the high spectro-spatial resolution of the ARKS data.

\section{Conclusions}
\label{sec:conclusion}

In this paper, we presented high spectrospatial resolution ALMA observations of $^{12}\textrm{CO}$ and $^{13}\textrm{CO}$ in the exoKuiper belt of HD~121617 as part of the large programme ARKS. Our main findings are as follows:

\begin{itemize}

    \item We presented linewidth and velocity integrated intensity maps, both showing an azimuthal X-shaped pattern, which we attribute to the azimuthal dependence of Keplerian shear, combined with high optical depth.
    \item We spectroscopically aligned and averaged spectra within annuli of $0\farcs02$ across the disc to create local spectral line profiles. We find that the local spectral line profiles across the disc are Gaussian in shape, with broad local linewidths where $^{12}\textrm{CO}$ displays a FWHM of $\sim1092$ m s$^{-1}$ at the peak radial surface density.
    \item We find that both optically thin and thick models can produce broad, Gaussian-shaped azimuthally averaged line profiles due to Keplerian shear and its azimuthal dependence; as such, the shape of these profiles alone cannot be used to differentiate between optically thin and thick emission.
    \item We modelled the local spectral line profiles with a simple radiative transfer toy model, neglecting Keplerian shear and find best-fit linewidth values that imply a kinetic temperature of hundreds of kelvin at the peak surface density if thermal broadening is dominant. Additionally, we find that our model implies that $^{12}\textrm{CO}$ and $^{13}\textrm{CO}$ are optically thin or marginally optically thin ($\tau\lesssim1$) at all radii.
    \item We find that a \texttt{RADMC-3D} optically thin model reproduces many key features of the data, including the spectral width (FWHM), peak intensity azimuthal dependence, and the 1D spectral line profile. However, the model can neither explain the X-shape seen in the velocity integrated intensity map nor the broader $^{12}\mathrm{CO}$ linewidth compared to $^{13}\textrm{CO}$. Thus, the optically thin model cannot explain our observations.
    \item We fitted the $^{13}\textrm{CO}$ data cube with a \texttt{RADMC-3D} model with a Gaussian CO surface density. We find that $^{13}\textrm{CO}$ is optically thick ($T_{\rm peak} = 38$ K, and $M_{\rm ^{13}\textrm{CO}}$ = $2 \times 10^{-3}$ M$_{\oplus}$) and has the same radial width as the dust. This model can reproduce all features of the data, including local linewidths and maps, and is therefore strongly preferred over optically thin models.
    \item We find that re-scaling this optically thick Keplerian $^{13}\textrm{CO}$ model by the ISM $^{12}\mathrm{CO}/^{13}\mathrm{CO}$ abundance ratio reproduces the bulk of the optically thick $^{12}\mathrm{CO}$ emission, while failing to reproduce the non-Keplerian nature of its velocity profile.

\end{itemize}

In conclusion, we find that both $^{12}\textrm{CO}$ and 
$^{13}\textrm{CO}$ in the HD~121617 gas-bearing debris disc are optically thick, confirming literature expectations. This implies that the derived CO mass is sensitive to model assumptions, and that we cannot derive kinetic temperatures and/or non-thermal broadening from the line profiles even when the effect of Keplerian shear is accounted for. With the above key caveats in mind, our model constrains the mean molecular weight to be $12.6_{-1.1}^{+1.3}$, which is likely through our model's measurement of the vertical location of the optically thick CO-emitting layer. This value of $\mu$ is significantly higher than if H$_{2}$ were the dominant gas species, suggesting a non-primordial composition. Overall, to break model degeneracies and achieve stronger constraints on fundamental gas properties such as its mass, temperature, and composition, future observations of optically thin lines at similar spectrospatial resolution will be necessary.

\begin{acknowledgements} 
We thank the anonymous reviewer for their careful review and for providing constructive comments that helped improve this paper. This paper makes use of the following ALMA data: ADS/JAO.ALMA 2022.1.00338.L, 2012.1.00142.S, 2012.1.00198.S, 2015.1.01260.S, 2016.1.00104.S, 2016.1.00195.S, 2016.1.00907.S, 2017.1.00167.S, 2017.1.00825.S, 2018.1.01222.S and 2019.1.00189.S. ALMA is a partnership of ESO (representing its member states), NSF (USA) and NINS (Japan), together with NRC (Canada), MOST and ASIAA (Taiwan), and KASI (Republic of Korea), in cooperation with the Republic of Chile. The Joint ALMA Observatory is operated by ESO, AUI/NRAO and NAOJ. The National Radio Astronomy Observatory is a facility of the National Science Foundation operated under cooperative agreement by Associated Universities, Inc. The project leading to this publication has received support from ORP, that is funded by the European Union’s Horizon 2020 research and innovation programme under grant agreement No 101004719 [ORP]. We are grateful for the help of the UK node of the European ARC in answering our questions and producing calibrated measurement sets. This research used the Canadian Advanced Network For Astronomy Research (CANFAR) operated in partnership by the Canadian Astronomy Data Centre and The Digital Research Alliance of Canada with support from the National Research Council of Canada the Canadian Space Agency, CANARIE and the Canadian Foundation for Innovation. AB acknowledges research support by the Irish Research Council under grant GOIPG/2022/1895. LM and SMM acknowledges funding by the European Union through the E-BEANS ERC project (grant number 100117693), and by the Irish research Council (IRC) under grant number IRCLA- 2022-3788. Views and opinions expressed are however those of the author(s) only and do not necessarily reflect those of the European Union or the European Research Council Executive Agency. Neither the European Union nor the granting authority can be held responsible for them. SM was supported by a Royal Society University Research Fellowship (URF- R1-221669).AMH acknowledges support from the National Science Foundation under Grant No. AST-2307920. PW acknowledges support from FONDECYT grant 3220399 and ANID -- Millennium Science Initiative Program -- Center Code NCN2024\_001. PM acknowledges research support by the National Science and Technology Council of Taiwan under grant NSTC 112-2112-M-001-032-MY3. Support for BZ was provided by The Brinson Foundation. A.A.S. is supported by the Heising-Simons Foundation through a 51 Pegasi b Fellowship. MB acknowledges funding from the Agence Nationale de la Recherche through the DDISK project (grant No. ANR-21-CE31-0015). EM acknowledges support from the NASA CT Space Grant. TDP is supported by a UKRI Stephen Hawking Fellowship and a Warwick Prize Fellowship, the latter made possible by a generous philanthropic donation. CdB acknowledges support from the Spanish Ministerio de Ciencia, Innovaci\'on y Universidades (MICIU) and the European Regional Development Fund (ERDF) under reference PID2023-153342NB-I00/10.13039/501100011033, from the Beatriz Galindo Senior Fellowship BG22/00166 funded by the MICIU, and the support from the Universidad de La Laguna (ULL) and the Consejer\'ia de Econom\'ia, Conocimiento y Empleo of the Gobierno de Canarias. SP acknowledges support from FONDECYT Regular 1231663 and ANID Millennium Science Initiative Program Center Code NCN2024\_001. JM acknowledges funding from the Agence Nationale de la Recherche through the DDISK project (grant No. ANR-21-CE31-0015) and from the PNP (French National Planetology Program) through the EPOPEE project. EC acknowledges support from NASA STScI grant HST-AR-16608.001-A and the Simons Foundation.
\end{acknowledgements} 

\section{Data availability}
The ARKS data used in this paper can be found in the \href{https://dataverse.harvard.edu/dataverse/arkslp}{ARKS dataverse}. For more information, visit \href{https://arkslp.org}{arkslp.org}.

\bibliographystyle{aa}
\bibliography{refs} 

\begin{appendix}

\section{Keplerian shear: Analytical formulation}
\label{sec:keplerianshear}

Previous studies have attempted to describe the effect of Keplerian shear on line broadening analytically; however, these discussions have often been incomplete or incorrect. For example, in \citet{Yen(2016)} (Eq. 7), the broadening due to Keplerian shear is estimated using an incorrect formula, specifically employing the mixed derivative \( \frac{d^2 v}{dr\, d\theta} \). This expression describes how the quantity \( \frac{dv}{d\theta} \) changes with radius \( r \), but it does not directly relate to the line broadening caused by beam convolution. Consequently, it does not provide a valid estimate of the shear-induced broadening. Furthermore, this formulation predicts zero broadening at \( \theta = 0 \), which is unphysical. This appendix presents a more accurate analytical treatment of Keplerian shear broadening.

In a circumstellar gas disc where the velocity field is Keplerian, or well approximated by Keplerian (neglecting pressure gradients and self-gravity), the azimuthal velocity in the orbital plane is defined by the balance of the gravitational and centrifugal force through the usual $v_{\phi} (r_{\rm orb}) = \sqrt{\frac{GM_{\ast}}{r_{\rm orb}}}$ (for an assumed circular orbit at distance $r_{\rm orb}$ from a star of mass $M_{\ast}$). For a planetary system inclined from face-on by an angle $i$, this leads to a projected line of sight velocity (leading to a measurable Doppler shift) of $v_{z\rm,sky}=\sqrt{\frac{GM_{\ast}}{r_{\rm orb}}}\cos(\phi_{\rm orb})\sin(i)$, where $\phi_{\rm orb}$ (measured on the plane of the disc) is the azimuth measured counter-clockwise from the orbital plane's $x_{\rm orb}$ axis. This axis corresponds to rotating about the axis inclined by the angle $i$ to transform from the orbital plane to the sky plane, resulting in $x_{\rm orb} = x_{\rm sky}$. Here, $x_{\rm sky}$ denotes the direction along the disc's semi-major axis in the sky image, which is itself rotated within the sky plane by the PA relative to north.

We here analytically treat geometrical and observational effects that have to be considered when analysing resolved spectral line profiles in Keplerian discs, affecting the observed line shapes in a way that is dependent on orbital radius $r_{\rm orb}$, azimuth $\phi_{\rm orb}$. We assume that the disc is vertically thin (2D) and therefore neglect effects caused by a disc's finite vertical structure, even though these can play a role as discussed using detailed 3D radiative transfer modelling (Section \ref{sec:RADMC-3D Modelling}).

Keplerian shear refers to the intrinsic property of a rotating disc where the orbital velocity varies with radius, with material closer to the star orbiting faster than material farther out. This creates significant radial and azimuthal velocity gradients within the disc plane. When observing such a disc, the line-of-sight component of these velocities is what we measure through Doppler shifts. Due to the finite spatial resolution of our observations, each measured velocity at a sky location ($x_{\rm sky}, y_{\rm sky}$) or ($r_{\rm sky}, \phi_{\rm sky}$) represents an average over a 2D area corresponding to the telescope’s synthesised beam. In regions where Keplerian shear produces strong velocity gradients within the beam, this averaging leads to beam smearing effects that must be accounted for.

The projected Keplerian velocity field can be thought of as a 2D scalar field $v_{z\rm,sky}(r_{\rm sky}, \phi_{\rm sky})$, and its maximum rate of change (projected acceleration) in any direction is represented by the magnitude of its gradient,
\begin{equation}
\|\nabla v_{z\rm,sky}\|=\sqrt{\left(\frac{\partial v_{z\rm,sky}}{\partial r_{\rm sky}}\right)^{2}+\left(\frac{1}{r_{\rm sky}}\frac{\partial v_{z\rm,sky}}{\partial \phi_{\rm sky}}\right)^{2}}.
\label{eq:gradmag}
\end{equation} 

To obtain this gradient analytically, we first have to rewrite $v_{z\rm,sky}$ as a function of sky, rather than orbital, coordinates, to go from
\begin{equation}
	v_{z\rm,sky}=\sqrt{\frac{GM_{\ast}}{r_{\rm orb}}}\sin i\cos\phi_{\rm orb}
    \label{eq:vzskyorb}
\end{equation}
to 
\begin{equation}
	v_{z\rm,sky}=\sqrt{\frac{GM_{\ast}}{r_{\rm sky}}}\sin i \ \frac{\mathrm{sgn}(\cos\phi_{\rm sky})}{\sqrt{\|\cos\phi_{\rm sky}\|}}\left(1+\frac{\tan^2\phi_{\rm sky}},{\cos^2i}\right)^{-\frac{3}{4}}
\end{equation}
using a change of coordinates from orbital to sky plane:
\begin{equation}
	\begin{cases}
		r_{\rm orb}=r_{\rm sky}\|{\cos\phi_{\rm sky}}\|\sqrt{1+\frac{\tan^2\phi_{\rm sky}}{\cos^2i}}\\
		\phi_{\rm orb}=\arctan\left(\frac{\tan{\phi_{\rm sky}}}{\cos{i}}\right)=\arccos\left[\mathrm{sgn}(\cos{\phi_{\rm sky}})\left(1+\frac{\tan^2\phi_{\rm sky}}{\cos^2i}\right)^{-\frac{1}{2}}\right].
		\label{eq:changecoords}
	\end{cases}
\end{equation}
We can then evaluate the partial derivatives in Eq. \ref{eq:gradmag} to obtain the magnitude of the local gradient in the radial and azimuthal sky directions,
{\small
\begin{equation}§
	\frac{\partial v_{z\rm,sky}}{\partial r_{\rm sky}}=-\frac{1}{2}\frac{v_{z\rm,sky}}{r_{\rm sky}}=-\ \frac{(GM_{\ast})^{1/2}}{2\ r_{\rm sky}^{3/2}}\sin i \ \frac{\mathrm{sgn}(\cos\phi_{\rm sky})}{\sqrt{\|\cos\phi_{\rm sky}\|}}\left(1+\frac{\tan^2\phi_{\rm sky}}{\cos^2i}\right)^{-\frac{3}{4}},
	\label{eq:rgrad}
\end{equation}
}
and
{\small
\begin{equation}
	\frac{1}{r_{\rm sky}}\frac{\partial v_{z\rm,sky}}{\partial \phi_{\rm sky}}= \frac{(GM_{\ast})^{1/2}}{2\ r_{\rm sky}^{3/2}}\sin i\ \frac{\sin \phi_{\rm sky}\left[\cos^2\phi_{\rm sky}\left(1+\frac{\tan^2\phi_{\rm sky}}{\cos^2i}\right)-\frac{3}{\cos^2 i}\right]}{(\|\cos\phi_{\rm sky}\|)^{7/2}\left(1+\frac{\tan^2\phi_{\rm sky}}{\cos^2i}\right)^{7/4}}.
	\label{eq:phigrad}
\end{equation}
}

Along the on-sky radial direction, both the radial and azimuthal gradients decrease in magnitude as the radius increases, as would be expected since the radial derivative of the Keplerian velocity itself decreases with distance from the star (projected or not). In fact, we expect the overall magnitude of the projected velocity gradient, and therefore of the Keplerian shear effect, to decrease in line with the $r^{-3/2}$ dependence above.

The radial and azimuthal gradients have a less straightforward dependence on sky azimuth $\phi_{\rm sky}$. In order to analyse it, we fix $r_{\rm orb}$ to 73 au as in our peak intensity annulus of HD~121617 and create an array of orbital azimuths $\phi_{\rm orb}$, calculate the respective ($r_{\rm sky}$, $\phi_{\rm sky}$) through Eq. \ref{eq:changecoords}, and input them into Eqs. \ref{eq:rgrad} and \ref{eq:phigrad} to obtain the gradients' azimuthal dependence. It is important to note that the behaviour of the radial and azimuthal gradients with respect to sky azimuth changes depending on whether it is plotted as a function of the sky azimuth $\phi_{\rm sky}$ itself or of the orbital azimuth $\phi_{\rm orb}$. For example, in the main body of this work, we refer to $\phi$ as the azimuth, but this is the orbital azimuth $\phi_{\rm orb}$ rather than the sky azimuth, as it has been de-projected to the orbital plane.

\begin{figure}[h]
\centering
  \includegraphics[width=1\linewidth]{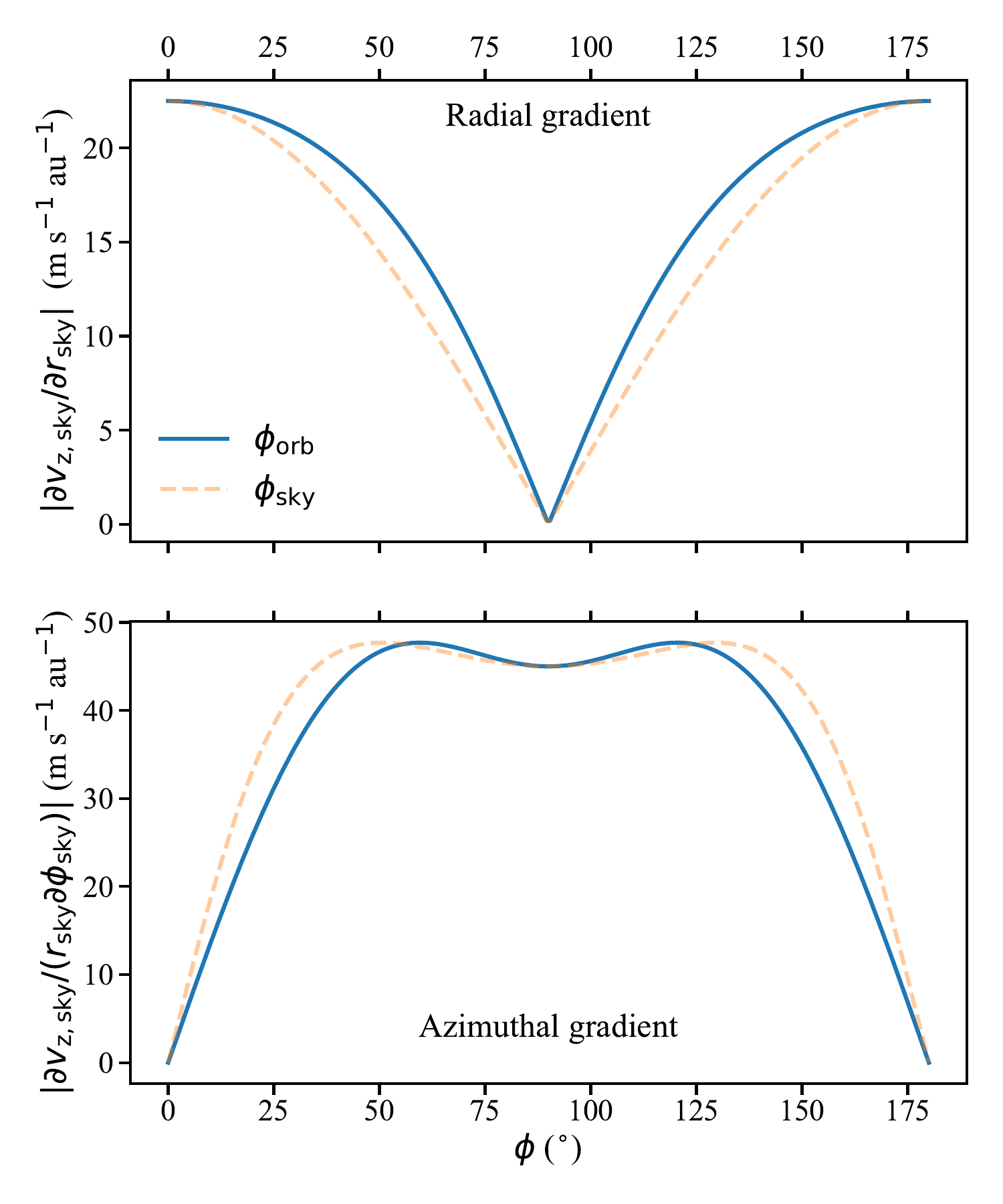}
  \caption{Azimuthal dependence of the radial and azimuthal gradient of the Keplerian projected line of sight velocity, providing an understanding of how the effect of Keplerian shear in the radial and azimuthal directions vary with azimuth. This is evaluated for a radius, stellar mass, and inclination representative of our ARKS target HD~121617. The solid blue and dashed orange lines show the dependence of the curve on whether orbital or sky azimuths are considered; the former should typically be considered for data that have been de-projected to the orbital plane.}
  \label{fig:gradients}
\end{figure}

Fig. \ref{fig:gradients} shows the absolute value of the azimuthal dependence of the radial and azimuthal gradients from Eqs. \ref{eq:rgrad} and \ref{eq:phigrad}. As expected, the radial gradient (top panel) is zero at the disc's on-sky semi-minor axis, where the line of sight Keplerian velocity is zero at all radii, and maximum along the disc's on-sky semi-major axis, where the projected Keplerian velocities are highest at all radii, producing a stronger radial gradient. 

The azimuthal gradient (bottom panel) has an opposite, though more complex, azimuthal dependence. It is zero at the semi-major axis, and reaches a maximum as it approaches the semi-minor axis, but then exhibits a dip at the semi-minor axis itself. The initial increase makes sense as the derivative of the line of sight velocity with respect to orbital azimuth (Eq. \ref{eq:vzskyorb}) is an increasing function of orbital azimuth going from semi-major to semi-minor axis. However, as we approach the semi-minor axis ($\phi=90^{\circ}$) this effect is counteracted by the fact that for a fixed sky azimuth swath $\Delta\phi_{\rm sky}$, the range of orbital $\Delta\phi_{\rm orb}$ subtended decreases, and with it so should the azimuthal gradient of the projected Keplerian velocity. This balance between the two effects causes the characteristic tooth shape in Fig. \ref{fig:gradients} (bottom).

\begin{figure}[h]
\centering
  \includegraphics[width=1\linewidth]{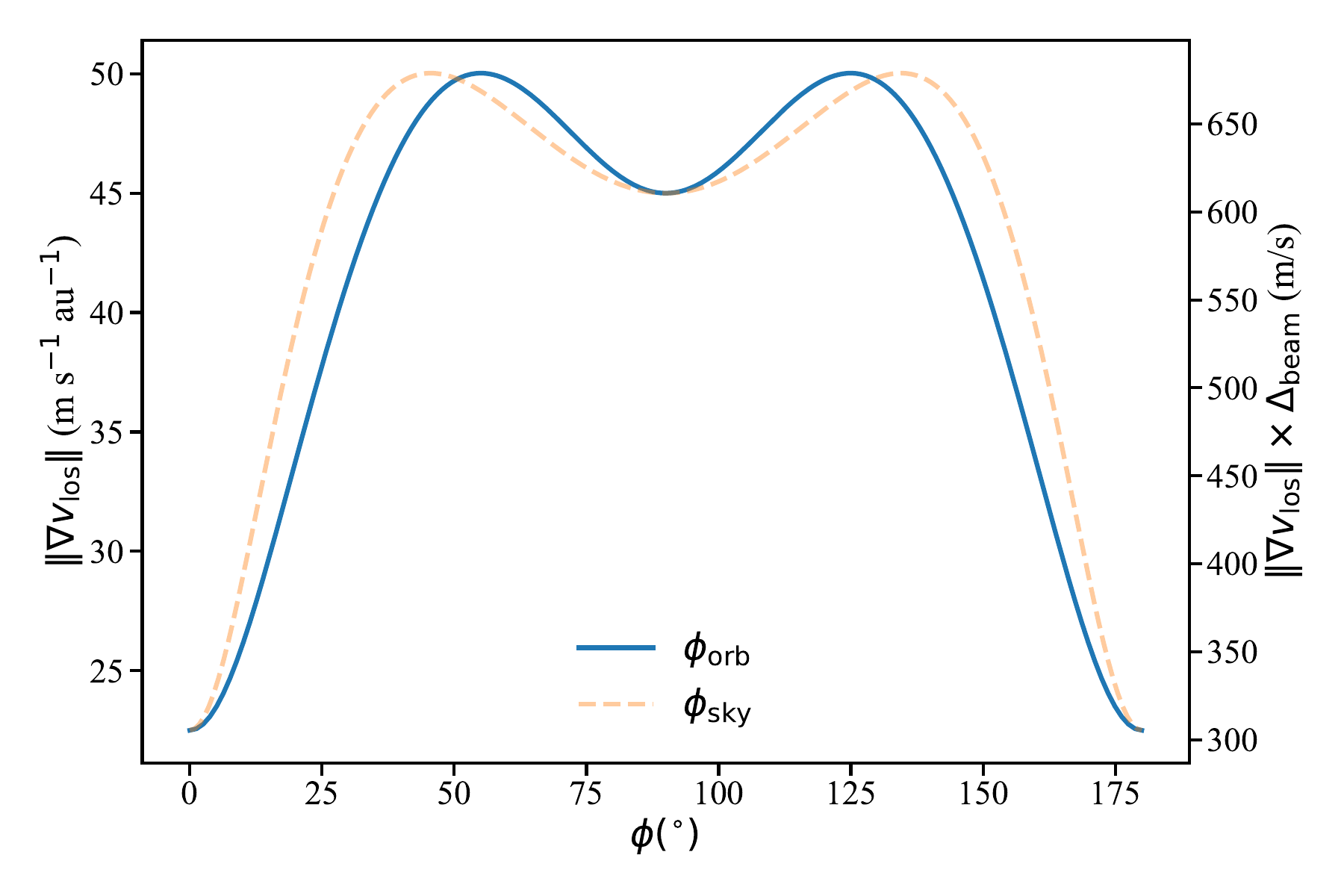}
  \caption{Azimuthal dependence of the effect of Keplerian shear, expressed as the maximum gradient of the Keplerian projected line-of-sight velocity in any direction (left y-axis), or evaluated as the same gradient by the beam size of our ARKS HD~121617 observations (right y-axis). The solid blue versus dashed orange lines show the dependence of the curve on whether orbital or sky azimuths are considered; the former should typically be considered for data that have been de-projected to the orbital plane.}
  \label{fig:analyticalshear}
\end{figure}

The magnitude of the maximum gradient in any direction, $\|\nabla v_{z\rm,sky}\|$ (Eq. \ref{eq:gradmag}), then shows the combination of the two effects as a function of azimuth in Fig. \ref{fig:analyticalshear}. This shows a more pronounced dip at the semi-minor axis but generally retains the same tooth-shaped pattern. This is in good qualitative agreement and explains the azimuthal behaviour of the linewidth seen in both our \texttt{RADMC-3D} models and our data (e.g. Fig. \ref{fig:FWHM}), as well as the X-shape seen in the linewidth and velocity integrated intensity maps (Fig. \ref{fig:12COmmaps}). 

This maximum gradient can be turned to a velocity difference expected due to Keplerian shear across a beam size $\Delta_{\rm beam}$ (in au), which will act to broaden local line profiles as discussed in the main text. This difference should, in principle, be evaluated through a line integral along the curve of the maximum gradient in the sky plane because the gradient's magnitude and direction change across the sky plane; however, we proceed for simplicity in the assumption that the gradient does not change significantly across a beam and evaluate this velocity difference by multiplying the gradient by the beam size in au (right y-axis in Fig. \ref{fig:analyticalshear}). In practice, this allows us to use our analytical calculations to estimate the magnitude of broadening by Keplerian shear for a given set of parameters ($M_{\star}, i, r_{\rm orb}, \phi_{\rm orb}, \Delta_{\rm beam}$). For the parameters of our ARKS HD~121617 observation, we find Keplerian shear should produce broadening of $\sim300-700$ m s$^{-1}$, with a strong (observed) azimuthal dependence.

\begin{figure}
\centering
  \includegraphics[width=1\linewidth]{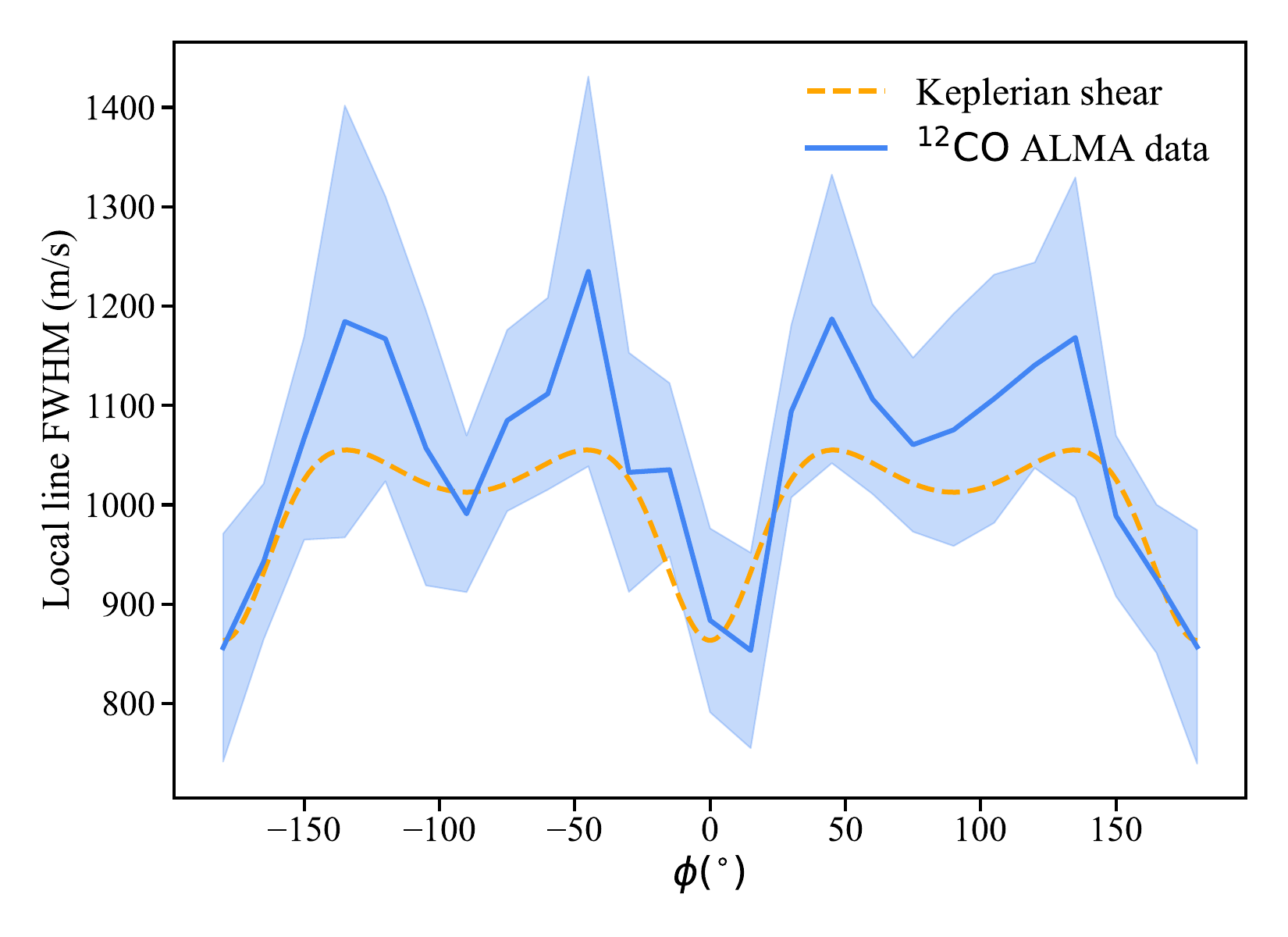}
  \caption{Azimuthal dependence of the local line profile, for direct comparison to the data in Fig. \ref{fig:FWHM}, obtained analytically through a simple combination of the Keplerian shear estimated in this section for HD~121617, the system's parameters at $r=73$ au, a broadened line profile for $^{12}$CO with a velocity width of $\sim 800$~m s$^{-1}$, and our instrumental spectral response with a resolution of 26 m s$^{-1}$.}
  \label{fig:analyticlocalprofiles}
\end{figure}

We can take this one step further to compare directly to our data under the simple assumption that the observed local line shape is simply a convolution of the line shape, a Gaussian with FWHM equal to the broadening due to Keplerian shear estimated here, and the spectral response of the instrument (in our case assumed to be Gaussian with FWHM of 26 m s$^{-1}$). We show the resulting line shape as a function of azimuth, assuming a broadened line profile similar to that derived in our \texttt{RADMC-3D} modelling ($\sigma_{v}$ $\sim 800~$m s$^{-1}$), in Fig. \ref{fig:analyticlocalprofiles}. 

This shows that our analytical model matches relatively well the FWHM linewidths of the local line profiles measured in our data and produced through \texttt{RADMC-3D} radiative transfer modelling in Fig. \ref{fig:FWHM}, with linewidths of $\sim$850-1050 m s$^{-1}$, and azimuthal peaks at $\sim\pm$55$^{\circ}$ and $\sim\pm$125$^{\circ}$. Some important caveats remain in this analysis and could explain the marginally higher peak linewidths in the observed azimuthal profile and/or deeper semi-minor axis dips; first, we did not consider vertical structure, which our \texttt{RADMC-3D} model already showed to play an important role in setting the shape of this azimuthal profile (Section \ref{sec:RADMC-3D Modelling}). Additionally, we simplified the effect of beam convolution by simply looking at the maximum gradient within a beam 'length' and neglected changes in velocity gradient within a beam size. Finally, we neglected non-Keplerian dynamics, which we know play a role in this system \citep{Seba_2arks}. Despite these caveats, we have shown that the effect of Keplerian shear on the local linewidths due to velocity gradients within spatial resolution elements can be estimated analytically and compares well with \texttt{RADMC-3D} models and ARKS observations of the HD~121617 system.  
\clearpage

\section{Additional figures}
\label{sec:addional_figures}

\begin{center}
\includegraphics[width=\linewidth]{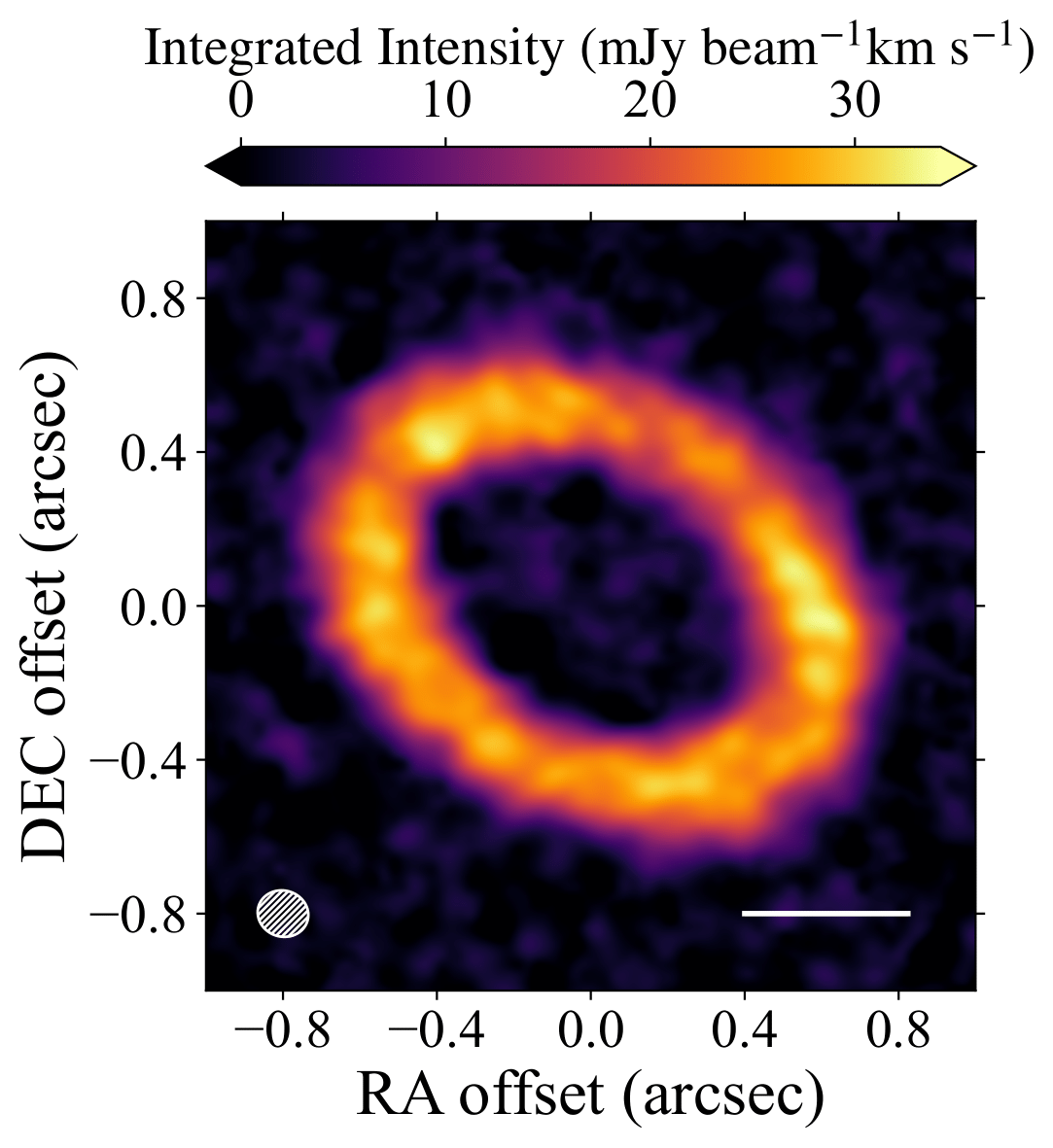}
\captionof{figure}{Velocity integrated intensity map for $^{13}\textrm{CO}$. The beam size is shown in the lower-left corner of each panel. The horizontal white bar indicates 50 au.}
\label{fig:13COmmaps}
\end{center}

\begin{center}
\includegraphics[width=0.95\linewidth]{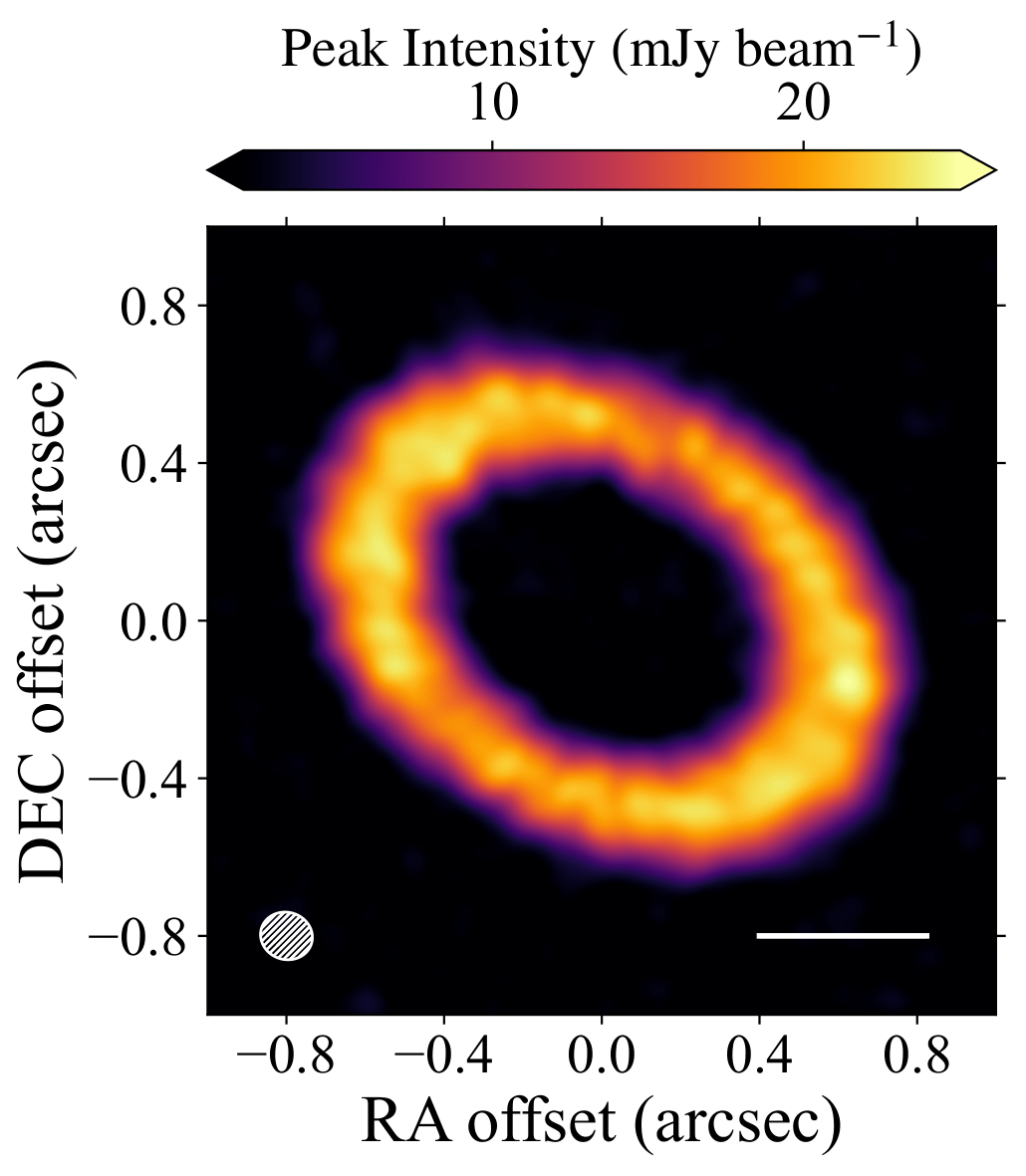}
\captionof{figure}{Peak intensity map for $^{13}\textrm{CO}$. The beam size is shown in the lower-left corner of each panel. The horizontal white bar indicates 50 au.}
\label{fig:13COmmaps_2}
\end{center}

\begin{figure}
\centering
  \includegraphics[width=0.8\linewidth]{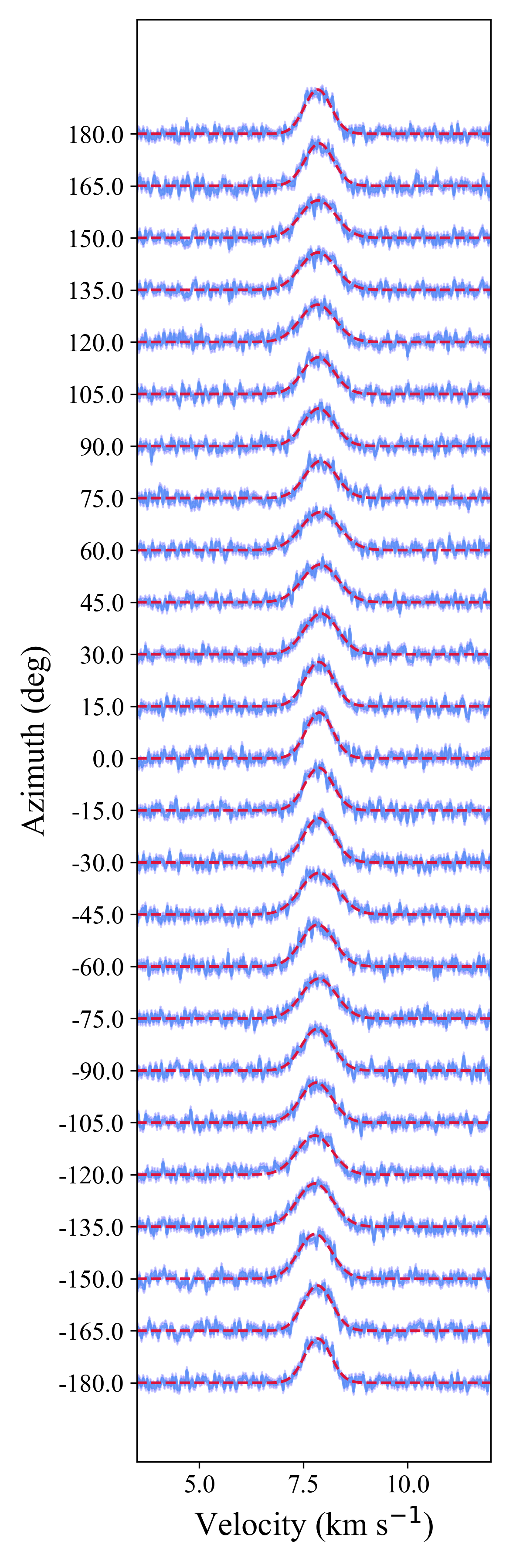}
  \caption{Line profiles (blue) for $^{12}\textrm{CO}$ extracted from an annulus ($0\farcs61$-$0\farcs63$) as a function of azimuth. Best-fit Gaussian models using an MCMC approach are overplotted (red).}
  \label{fig:stacked_azimuth_river}
\end{figure}

\begin{figure}
\centering
  \includegraphics[width=\linewidth]{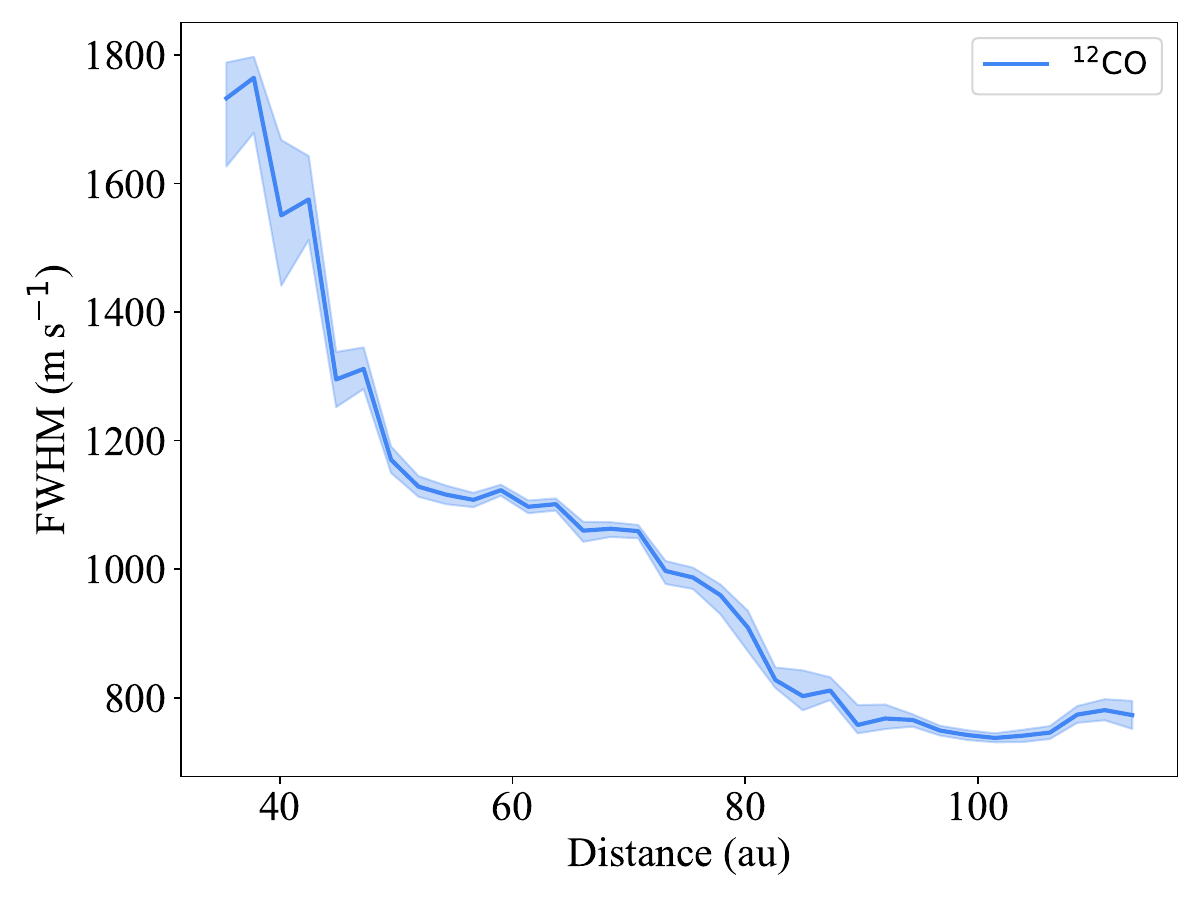}
  \caption{Best-fit value for $^{12}\textrm{CO}$ FWHM linewidth calculated from the kinetic temperature as a function of distance. 
}
  \label{fig:best_params}
\end{figure}

\begin{figure}
\centering
  \includegraphics[width=\linewidth]{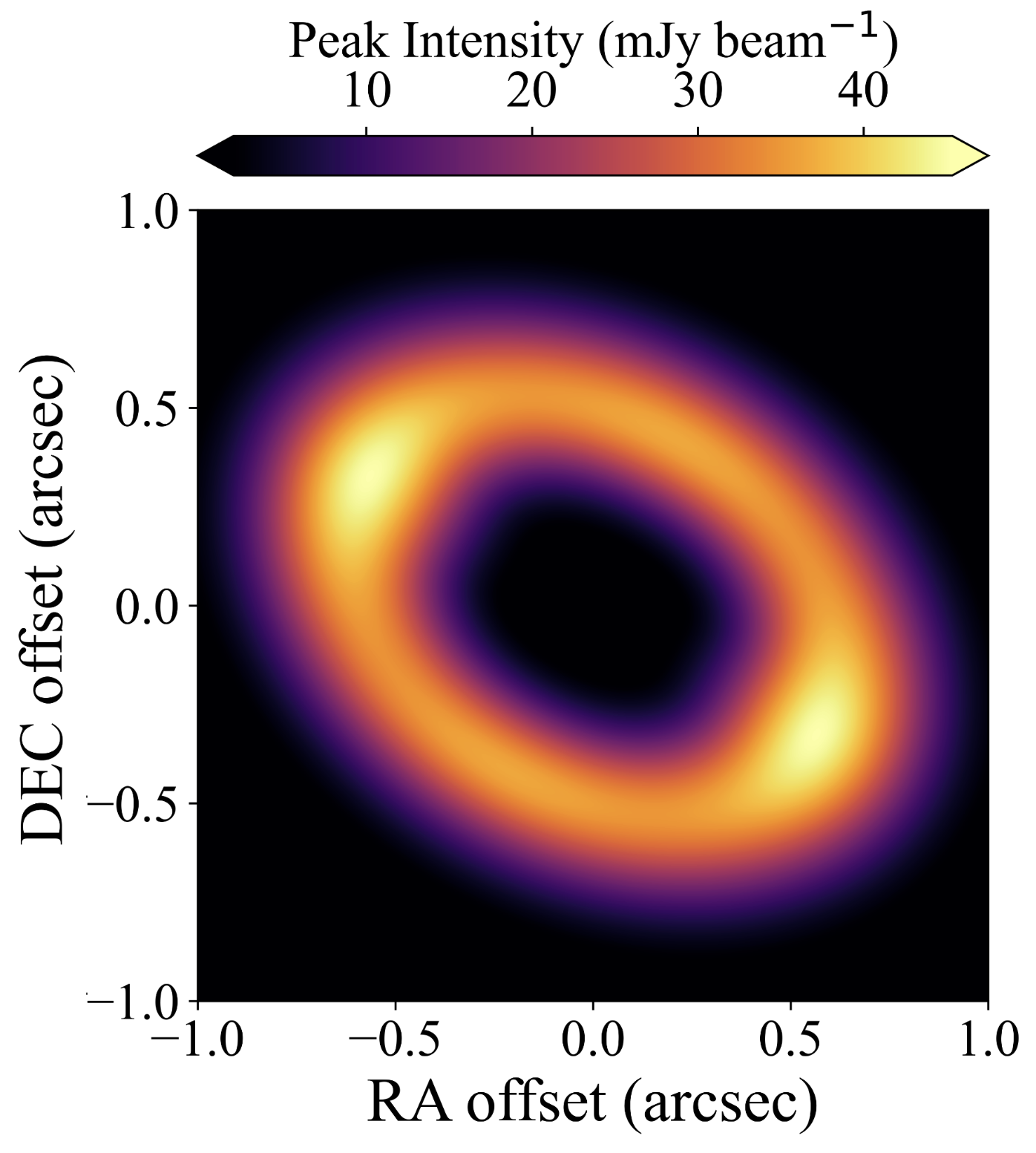}
  \caption{Peak intensity map for $^{12}\textrm{CO}$ optical thin \texttt{RADMC-3D} model presented in Section \ref{sec:Optically Thin Model}.}
  \label{fig:opticalthin_mom8}
\end{figure}

\begin{figure}
\centering
  \includegraphics[width=\linewidth]{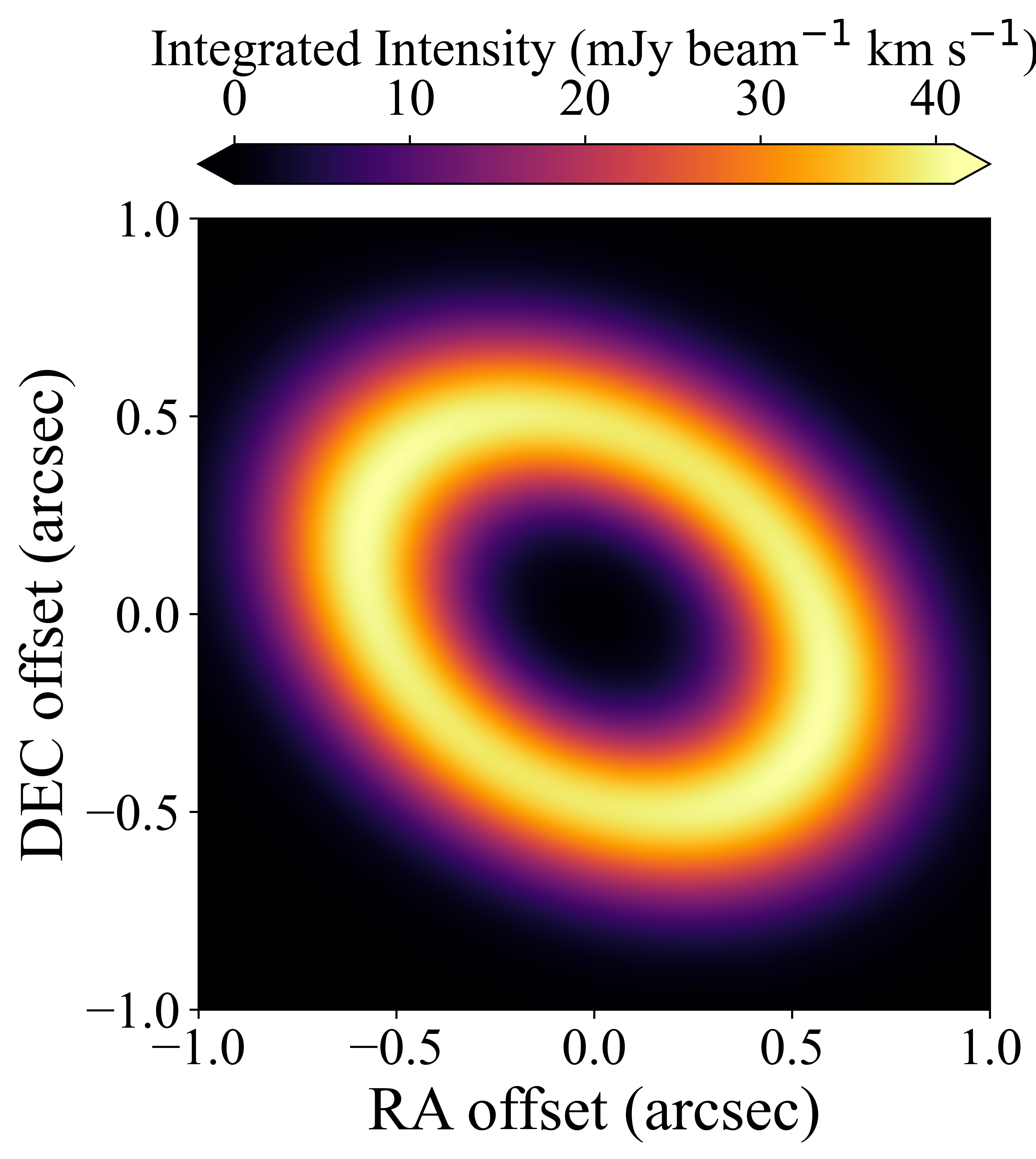}
  \caption{Velocity integrated intensity map for $^{12}\textrm{CO}$ optical thin \texttt{RADMC-3D} model presented in Section~\ref{sec:Optically Thin Model}.}
  \label{fig:opticalthin_mom0}
\end{figure}

\begin{figure*}
\centering
  \includegraphics[width=\linewidth]{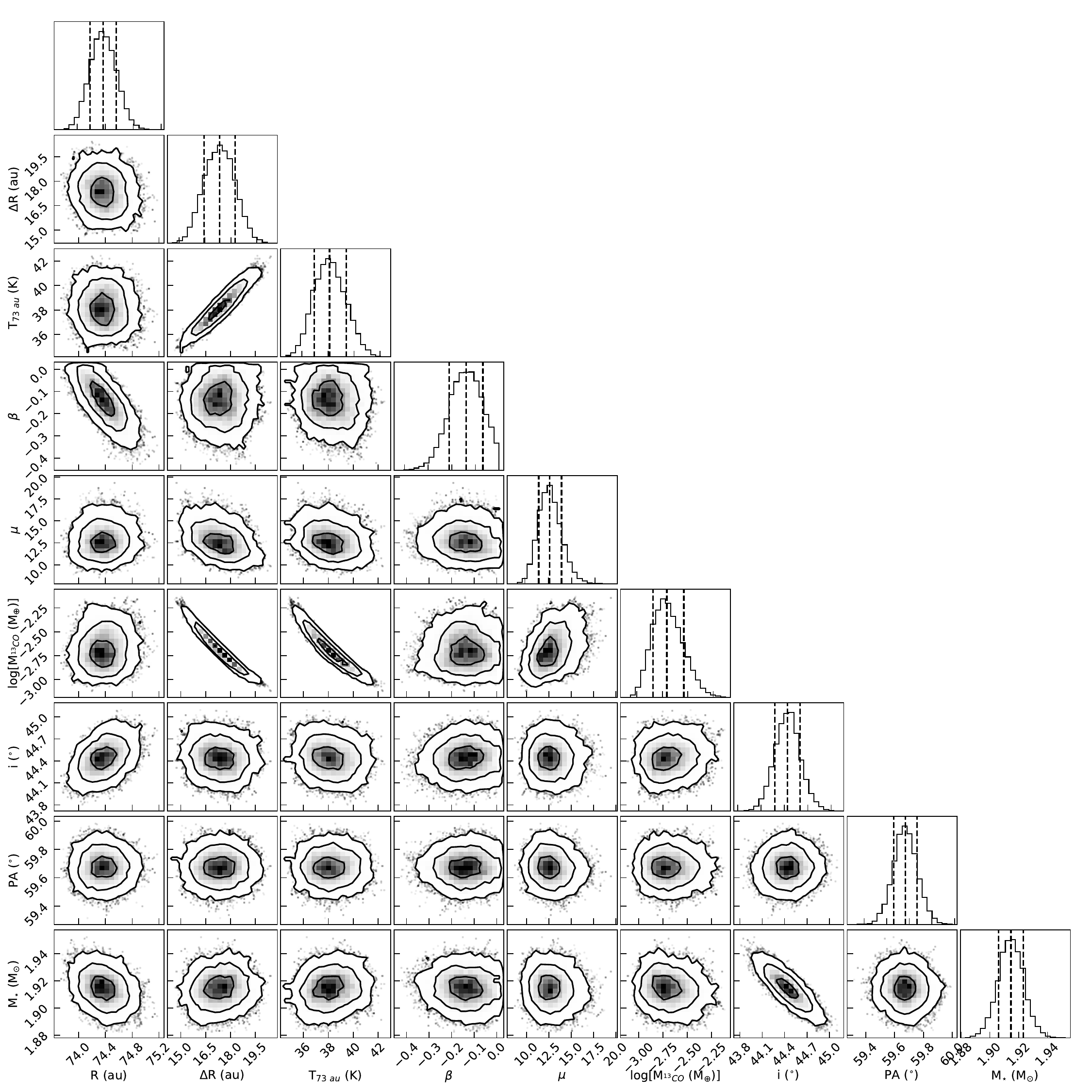}
  \caption{Posterior distribution of the $^{13}\textrm{CO}$ model. $R$ (au) is the peak of the Gaussian surface density distribution; $\Delta R$ (au), the FWHM of the Gaussian distribution; $T_{73}$ (K), the temperature at 73 au; $\beta$, the temperature power-law index; $\mu$, the mean molecular weight; $M_{CO}$ (M$_{\oplus}$), the CO mass; $i$ (deg), the inclination; $\mathrm{PA}$ (deg), the position angle; and $M_{\star}$ (M$_{\odot}$), the stellar mass. The marginalised distributions are presented in the diagonal. The best-fitting values and uncertainties for the parameters are taken from the 16th, 50th, and 84th percentiles (vertical dashed lines).}
  \label{fig:13CO_gauss_corner}
\end{figure*}

\begin{figure*}
\centering
  \includegraphics[width=0.8\linewidth]{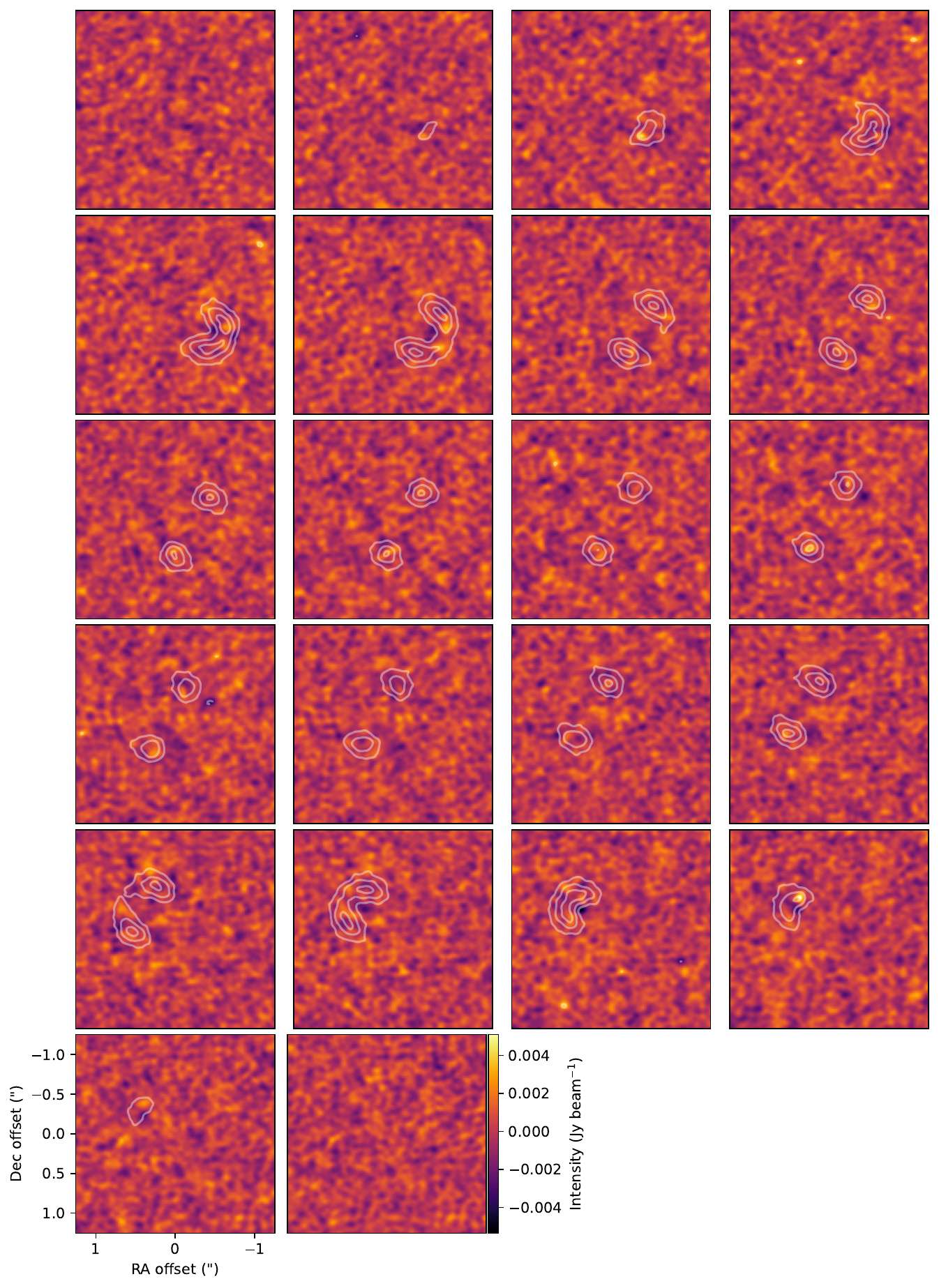}
  \caption{Residual channel maps for $^{13}\textrm{CO}$ showing $^{13}\textrm{CO}$ data minus the best-fit $^{13}\textrm{CO}$ model. White contours indicate the position of the disc in each channel.}
  \label{fig:13COchannelmap}
\end{figure*}

\begin{figure*}
\centering
  \includegraphics[width=0.8\linewidth]{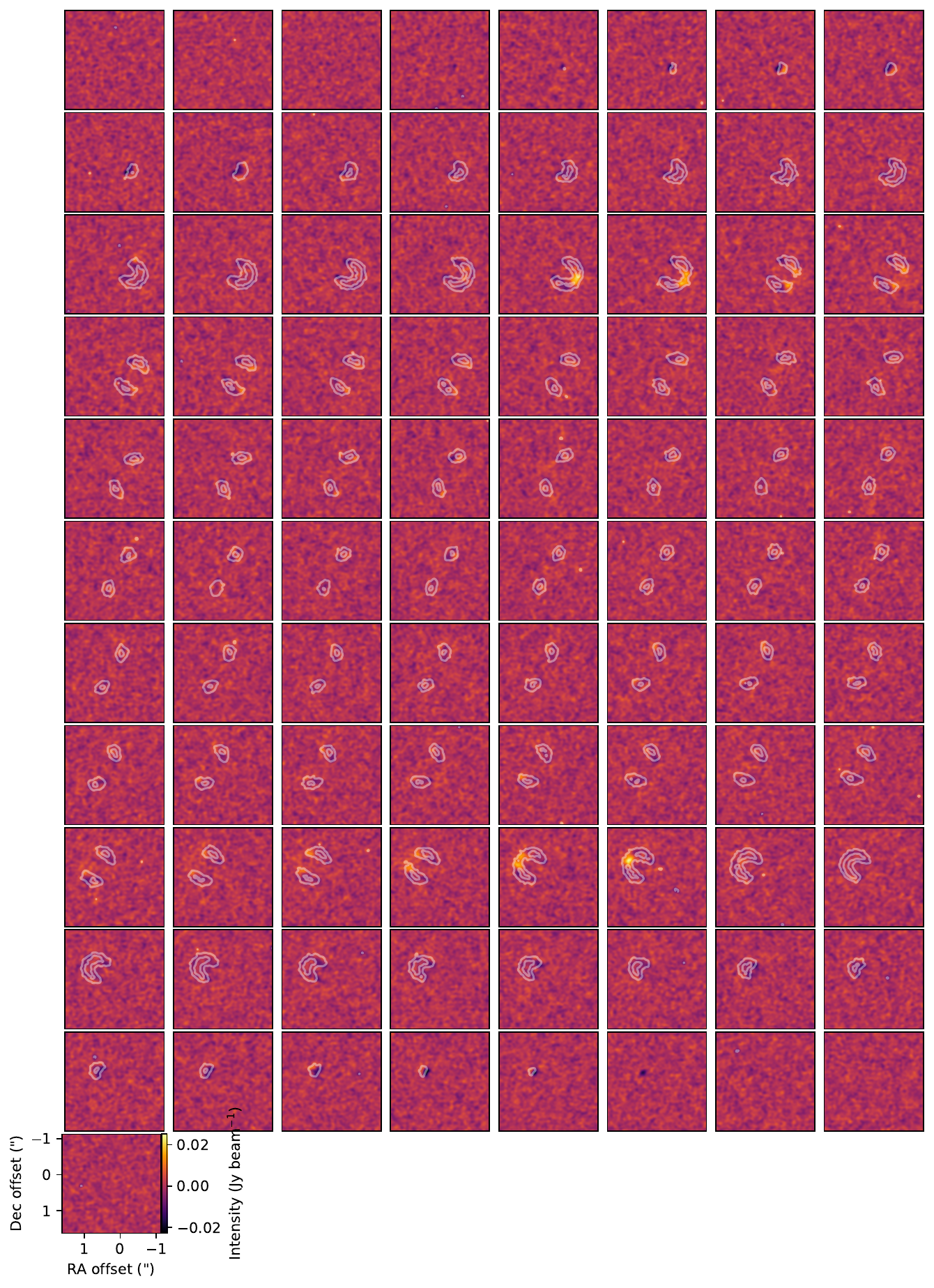}
  \caption{Residual channel maps for $^{12}\textrm{CO}$ showing $^{12}\textrm{CO}$ data minus the $^{12}\textrm{CO}$ model produced by re-scaling the best-fit $^{13}\textrm{CO}$ model by the ISM abundance ratio. Additionally, white contours indicate the positions of the disc in each channel.}
  \label{fig:12COchannelmap}
\end{figure*}

\end{appendix}
\end{document}